\documentclass{aa}

\usepackage{graphicx}
\usepackage{txfonts}
\usepackage{hyperref}
\hypersetup{
    colorlinks=true,
    linkcolor=blue,
    filecolor=blue,      
    urlcolor=blue,
    citecolor=blue,
    pdftitle={Lens parameters for Gaia18cbf -- a long gravitational microlensing event in the Galactic plane}
    }

\usepackage{multirow}

\defcitealias{MW}{MW}
\defcitealias{2013PecautMamajek}{PM}

\newcommand{\tnod}{t_\mathrm{0}}
\newcommand{\unod}{u_\mathrm{0}}
\newcommand{\te}{t_\mathrm{E}}
\newcommand{\pien}{\pi_\mathrm{EN}}
\newcommand{\piee}{\pi_\mathrm{EE}}
\newcommand{\pie}{\pi_\mathrm{E}}
\newcommand{\Gnod}{G_\mathrm{0}}
\newcommand{\fsg}{f_{\mathrm{S},G}}
\newcommand{\inod}{i_\mathrm{0}}
\newcommand{\fsi}{f_{\mathrm{S},i}}
\newcommand{\Fs}{F_{\mathrm{s}}}
\newcommand{\Fb}{F_{\mathrm{b}}}
\newcommand{\thetae}{\theta_\mathrm{E}}

\newcommand{\dl}{D_\mathrm{L}}
\newcommand{\ml}{M_\mathrm{L}}
\newcommand{\ds}{D_\mathrm{S}}

\newcommand{\Gbl}{G_{\mathrm{blend}}}

\titlerunning{Lens parameters for Gaia18cbf} 
\authorrunning{Kruszy\'nska et al.}

\begin{document}

   \title{Lens parameters for Gaia18cbf -- a long gravitational microlensing event in the Galactic plane\thanks{Tables \ref{tab:photGaia} and \ref{tab:photGaiaCSPSi} are only available in electronic form at the CDS via anonymous ftp to cdsarc.u-strasbg.fr (130.79.128.5) or via http://cdsweb.u-strasbg.fr/cgi-bin/qcat?J/A+A/.}}
   
   \author{Katarzyna Kruszy\'nska\inst{\ref{oauw}}
        \and
        {\L{}}.~Wyrzykowski\inst{\ref{oauw}}
        \and
            K.~A.~Rybicki\inst{\ref{oauw}}
            \and
            M.~Maskoli\={u}nas\inst{\ref{vilnius}}
            \and 
            E.~Bachelet\inst{\ref{lco}}
            \and
            N.~Rattenbury\inst{\ref{auck}}
            \and
            P.~Mr{\'o}z\inst{\ref{oauw}}
            \and
                P.~Zieli{\'n}ski\inst{\ref{iaumk},\ref{oauw}}
                \and
        K.~Howil\inst{\ref{oauw}}
        \and
        Z.~Kaczmarek\inst{\ref{oauw}, \ref{ioa}}
                \and
                S.~T.~Hodgkin\inst{\ref{ioa}}
            \and
        N.~Ihanec\inst{\ref{oauw}}
        \and
        I.~Gezer\inst{\ref{oauw}}
        \and
        M.~Gromadzki\inst{\ref{oauw}}
            \and 
            P.~Miko{\l{}}ajczyk\inst{\ref{oauw},\ref{uwr}}
        \and
        A.~Stankevi\v{c}i\={u}t\.{e}\inst{\ref{oauw}}
        \and
        V.~\v{C}epas\inst{\ref{vilnius}}
        \and
        E.~Pak\v{s}tien\.{e}\inst{\ref{vilnius}}
        \and 
        K.~\v{S}i\v{s}kauskait\.{e}\inst{\ref{vilnius}}
        \and
        J.~Zdanavi\v{c}ius\inst{\ref{vilnius}}
        \and
        V.~Bozza\inst{\ref{salerno},\ref{infn}}
        \and
        M.~Dominik\inst{\ref{standrews}}
        \and
        R.~Figuera~Jaimes\inst{\ref{rojofija}}
        \and
        A.~Fukui\inst{\ref{toudai}, \ref{canary}}
        \and
        M.~Hundertmark\inst{\ref{heidelberg}}
        \and 
        N.~Narita\inst{\ref{toudai}, \ref{canary}, \ref{astrobio}}
        \and
        R.~Street\inst{\ref{lco}}
        \and
        Y.~Tsapras\inst{\ref{heidelberg}}
        \and
        M.~Bronikowski\inst{\ref{oauw}, \ref{cacung}}
        \and
        M.~Jab{\l}o{\'n}ska\inst{\ref{oauw}}
        \and
        A.~Jab{\l}onowska\inst{\ref{oauw}}
        \and
        O.~Zi{\'o}{\l}kowska\inst{\ref{oauw}, \ref{camk}}
        }

   \institute{Astronomical Observatory, University of Warsaw, Al. Ujazdowskie 4, 00-478 Warszawa, Poland\\
              \email{kkruszynska@astrouw.edu.pl}
              \label{oauw}
              \and
              Institute of Theoretical Physics and Astronomy, Vilnius University, Saulėtekio al. 3, Vilnius, LT-10257, Lithuania
              \label{vilnius}
              \and
              Las Cumbres Observatory, 6740 Cortona Drive, suite 102, Goleta, CA 93117, USA
              \label{lco}
              \and
              Department of Physics, University of Auckland, Private Bag 92019, Auckland, New Zealand
              \label{auck}
              \and
              Institute of Astronomy, Faculty of Physics, Astronomy and Informatics, Nicolaus Copernicus University in Toru{\'n}, ul. Grudzi\k{a}dzka 5, 87-100 Toru{\'n}, Poland
              \label{iaumk}
              \and 
              Institute of Astronomy, University of Cambridge, Madingley Road, Cambridge, CB3 0HA, United Kingdom
              \label{ioa}
              \and
              Instytut Astronomiczny, Uniwersytet Wroc{\l}awski,  Kopernika 11, Wroc{\l}aw, Poland
              \label{uwr}
              \and
              Dipartimento di Fisica ``E.R. Caianiello'', Universit\`a di Salerno, Via Giovanni Paolo II, 84084, Fisciano, Italy
              \label{salerno}
              \and
              Istituto Nazionale di Fisica Nucleare, Sezione di Napoli, Via Cintia, 80126, Napoli, Italy
              \label{infn}
              \and
              University of St Andrews, Centre for Exoplanet Science, SUPA School of Physics \& Astronomy, North Haugh, St Andrews, KY16~9SS, United Kingdom
              \label{standrews}
              \and
              Facultad de Ingeniería y Tecnología, Universidad San Sebastian, General Lagos 1163, Valdivia 5110693, Chile
              \label{rojofija}
              \and
              Komaba Institute for Science, The University of Tokyo, 3-8-1 Komaba, Meguro, Tokyo 153-8902, Japan
              \label{toudai}
              \and
              Instituto de Astrof\'isica de Canarias, V\'ia L\'actea s/n, E-38205 La Laguna, Tenerife, Spain
              \label{canary}
              \and
              Zentrum f{\"u}r Astronomie der Universit{\"a}t Heidelberg, Astronomisches Rechen-Institut, M{\"o}nchhofstr. 12-14, 69120 Heidelberg, Germany
              \label{heidelberg}
              \and
              Astrobiology Center, 2-21-1 Osawa, Mitaka, Tokyo 181-8588, Japan
              \label{astrobio}
              \and
              Nicolaus Copernicus Astronomical Center, Polish Academy of Sciences, Bartycka 18, PL-00-716 Warsaw, Poland
              \label{camk}
              \and
              Center for Astrophysics and Cosmology, University of Nova Gorica, Vipavska 11c, SI-5270 Ajdovščina, Slovenia
              \label{cacung}
             }

   \date{Received date /
Accepted date }

 
  \abstract
{The  timescale  of  a  microlensing  event  scales  as  a  square  root  of  a  lens  mass.  Therefore,   
long-lasting  events  are  important  candidates  for  massive  lenses,  including  black  holes. 
}   
   {Here, we present the analysis of the Gaia18cbf microlensing event reported by the Gaia Science Alerts system. It exhibited a long timescale and features that are common for the annual microlensing parallax effect. We deduce the parameters of the lens based on the derived best fitting model. }
   {We used photometric data collected by  the Gaia satellite as well as the follow-up data gathered by the ground-based observatories. We investigated the range of microlensing models and used them to derive the most probable mass and distance to the lens using a Galactic model as a prior. Using a known mass-brightness relation, we determined how likely it is that the lens is a main-sequence (MS) star.}
   {This event is one of the longest ever detected, with the Einstein timescale of $t_\mathrm{E}=491.41^{+128.31}_{-84.94}$~days for the best solution and $t_\mathrm{E}=453.74^{+178.69}_{-105.74}$~days for the second best. Assuming Galaxy priors, this translates to the most probable lens masses of $\ml = 2.65^{+5.09}_{-1.48} M_\odot$ and $\ml = 1.71^{+3.78}_{-1.06} M_\odot$, respectively. 
   The limits on the blended light suggest that this event was most likely not caused by a MS star, but rather by a dark remnant of stellar evolution.}
   {}

   \keywords{Gravitational lensing: micro -- Techniques: photometric -- white dwarfs -- Stars: neutron -- Stars: black holes --
               }

   \maketitle
%
\section{Introduction}


The first person to realise the possibility of gravitational microlensing was R. W. Mandl, who then inspired Einstein to make calculations that proved a technical possibility of such effect. Einstein himself, however, dismissed it as impossible to observe due to the rareness of such events and the limitations of the instruments present at that time \citep{1936Einstein}. 
During the 1960s, much of theoretical groundwork was done \citep{1964Liebes, 1964Refsdal}, but this phenomenon was not pursued as it is today until the idea was re-introduced towards the end of 1980 by Bohdan Paczy{\'n}ski as a way to reveal the nature of dark matter \citep{1986Paczynski}. This has launched many successful surveys including MACHO \citep{1992ASPC...34..193A, 1993NYASA.688..612B}, OGLE \citep{1992AcA....42..253U}, EROS \citep{1993Msngr..72...20A}, and MOA \citep{1997vsar.conf...75A} and led to a detection of many microlensing events. Some of the events revealed planets around microlensing stars, which was predicted by \cite{1991ApJ...374L..37M}. In order to obtain better light curves for observed events, follow-up with other telescopes was frequently used, including telescope networks such as 
$\mu$FUN \citep{2010microFun} or Las Cumbres Observatory network of robotic telescopes \citep{LCO}.
Recently, a new era of microlensing surveys started with the launch of KMTNet, which provides near 24 hour monitoring of selected fields towards the Galactic bulge with cadences from 15 minutes up to 5 hours \citep{2016JKAS...49...37K}. 

Despite astonishing improvements in the way data are acquired, it is often hard to measure physical parameters of microlensing system, 
because  it  is  difficult  to  uniquely  determine  the  mass  and  distance  to  the  lens  based only  on  the  event  timescale.
Information about the mass of the lens is encoded in a parameter called the Einstein radius, which is defined as
\begin{equation}
    \centering
    \theta_E = 0.9 mas \sqrt{\frac{\ml}{M_\odot}} \sqrt{\frac{10 \, \text{kpc} \,\, \ds}{\dl \,(\ds - \dl)}} \,\,,
\end{equation}
where $M_\mathrm{L}$ is the mass of the lens, $D_\mathrm{S}$ is the distance to the source, and $D_\mathrm{L}$ is the distance to the lens \citep{1996ARA&A..34..419P}.

Deriving the absolute mass of the lens is, however, almost impossible without second-order effects or additional data. 
One such effect is the finite-source effect, 
which has to be included when the source is crossing over the lens or caustic or in other cases, when the gradient of lensing  magnification over the source surface plays a significant role.
One example where this occurs is well-sampled high magnification events or  caustic-crossing events \citep{1994ApJ...430..505W}. 

Another method that can be used to obtain additional information about the lens system is parallax, which can be broken into three sub-groups: annual, terrestrial, and space. 
Annual parallax occurs when the event is long enough for Earth's movement around the Sun to change the position of the observer and produces a distinct asymmetry in the observed light curve \citep{1992Gould, 1995ApJ...454L.125A}.
Such an effect, combined with a large timescale $t_\mathrm{E}$ can be used to constrain the mass of the lens and to find microlensing dark remnant candidates \citep{2002Bennett, 2005ApJ...633..914P, 2016WyrzykBH, 2020WyrzykowskiMandelBH, 2020LamPopSyCLE, 2020KarolinskiZhu, 2020Abrams, MW}.
The longest ever detected microlensing event had a timescale of $640^{+68}_{-54}$~days \citep{2002MaoLongUlens}, and only a few known events have timescales around 400~days \citep{2015Wyrzyk, 2020MrozDisk}.
A terrestrial parallax occurs when a microlensing event is observed from two separate observatories on Earth. 
If the separation between the sites is large enough and the event is observed at high cadence, it is possible to see that the time of the peak and impact parameter differs between sites \citep{1995MNRAS.276L..79H, 1996ApJ...471...64H, 2009ApJ...698L.147G}. 
The space parallax is similar to the terrestrial one, but one of the observatories is located in space \citep{1966MNRAS.134..315R}. 
If the separation is as large as 1~au, it is possible to observe significant differences in the impact parameter $u_\mathrm{0}$ and the time of peak $t_\mathrm{0}$ even for standard events.
Such observations have been provided for many events observed simultaneously by OGLE and Spitzer Space Telescope, most notably \cite{2007ApJ...664..862D}.
If the event is binary and both observatories manage to catch a caustic crossing, the separation may be smaller: as low as ~0.01~au \citep{2020Gaia16aye}.
If it is possible to measure $\theta_\mathrm{E}$ and $\pi_\mathrm{E}$, for example through finite-source effect and a parallax effect, one can directly measure the mass of the lens using the following relation \citep{2002An, 2004Gould, RybickiGaia19bld}:
\begin{equation}
\label{eq:mass}
M = \frac{\thetae}{\kappa \pie},
\end{equation}
where $\kappa = \frac{4G}{c^2 \mathrm{au}} \approx 8.144 \frac{\mathrm{mas}}{M_\odot}$ \citep{2000ApJ...542..785G}.

A different approach to obtaining the mass of the lens is to measure the flux of the lens $F_\mathrm{L}$ \citep{2017Koshimoto, 2020Koshimoto}. 
This can be done with high angular resolution imaging using adaptive optics or the Hubble Space Telescope and requires knowledge of the mass-luminosity relation for the pass-band in which the event is observed. 
Such relations are known for the stars in the Galactic bulge, but not necessarily for the Galactic disc. 

Yet another way of obtaining additional data needed to break degeneracies of the microlensing model 
is to observe how the event affects the astrometric position of the source.
When the source is approaching the Einstein radius, its image gets deformed and splits into two separate images.
Because the separation between two images is too small for events occurring in the Milky Way to be observable by the majority of modern telescopes, we observe a shift of the centroid of light instead.
With enough precision, such effects can be measured, and it provides data for the direct measurement of the Einstein radius. 
So far, only one team has been successful in measuring an astrometric microlensing effect using the Hubble Space Telescope, and they revealed a microlensing event caused by a white dwarf \citep{2017Sci...356.1046S}.
In recent years, using the Very Large Telescope's GRAVITY or PIONIER has helped to resolve two deformed images of the microlensed source, which provided an addidional way to measure $\theta_\mathrm{E}$ \citep{2019DongVLTI, CassanGaia19bld}.


Gaia is a space satellite that was launched on 19 December 2019 from French Guiana by the European Space Agency. 
As a successor of the \textit{Hipparcos} mission, its main purpose is to measure positions and parallaxes of 1 billion stars residing in the Milky Way with sub-milliarcsecond precision. 
It is also going to measure space velocities of observed stars and create a 3D structural map of the Milky Way \citep{2016GaiaMission}. 
It is equipped with two telescopes sharing a common focal plane. 
These telescopes have curved (1.45x0.5)m mirrors and their fields of view are separated by an angle of approximately $106.5^\circ$. 
The focal plane consists of 104 CCD matrices, which include two rows of Sky Mapper CCDs responsible for detecting observable objects, nine rows of astro-photometric fields (AF) measuring the brightness of a source in a wide $G$ band, two rows of low-resolution spectro-photometers (BP and RP), and three rows of high-precision radial velocity spectrograph used for sources brighter than $V = 16$~mag (RVS). 
The spacecraft is currently residing on a Lissajous orbit around the Lagrangian L2 point of the Earth-Sun system. 
It is rotating along its axis with a period of six hours, while the rotational axis is precessing with a period of 63 days. 
This allows Gaia to make full scans of the Milky Way repeatedly during the five years of its nominal mission \citep{2016GaiaMission}. 
Such a design yields the following consequences for the observations of a single source. 
First of all, each time Gaia passes a field, where an interesting object resides, we obtain two scans separated by a few hours.
Secondly, instead of taking an image of the field, a source passing through a single row of CCDs is detected by the Sky Mappers, and then nine measurements of brightness are taken by AFs with a 3 s cadence -- this means that the blend flux observed by Gaia may be smaller than in the case of a ground-based telescope.
Lastly, on average, each source is observed once per 30 days -- this means that ground-based follow-up is necessary for an evenly covered light curve suitable for modelling.

Since Gaia is observing the whole sky, the idea of an alerting system, called Gaia Science Alerts (GSA)\footnote{\href{http://gsaweb.ast.cam.ac.uk/alerts/}{http://gsaweb.ast.cam.ac.uk/alerts/}}, was conceived \citep{2012IAUS..285..425W, 2013RSPTA.37120239H, 2021Hodgkin}. 
Gaia was expected to deliver not only astrometry with precision down to 24 microarcseconds, but also to provide a vast sample of interesting transients \citep{2016MNRAS.455..603B}. 
Gaia Science Alerts has been operating since 2014 and has revealed around 18000 transients, of which the majority are supernovae, active galactic nuclei, cataclysmic variables, and flaring young stellar objects.
Microlensing events are also detected, and to date more than 300 candidates have been identified, with the vast majority located in the Galactic plane. 
Because of sparse Gaia sampling, global follow-up campaigns have been organised and provide supplementary photometric data, which then help one to study the events in detail (for example \cite{2020Gaia16aye, RybickiGaia19bld}).
The astrometric capabilities of the Gaia mission are going to be particularly useful for hundreds of events observed by Gaia, as the sub-milliarcsecond precision for the measurement of the position of observed source will lead to the measurement of
the Einstein radius, which helps to determine the lens mass with improved precision
 \citep{2002MNRAS.331..649B, 2018MNRAS.476.2013R}. 




Here, we present a study of the photometric data of one of the longest microlensing events detected by the Gaia Science Alerts system, Gaia18cbf. 
We present a detailed analysis of this event and derive a probable range of parameters of the object that caused it.
This paper is organised as follows. In Section \ref{sec:discovery}, we present when and how the event was discovered and how photometric data were gathered.
Next, we describe microlensing model we used and how we found best fitting models in Section \ref{sec:model}.
Section \ref{sec:source_star} is devoted to discussing the possible location and nature of the source star, while in Section \ref{sec:lens} we discuss the nature of the lens. 
We discuss the results of the entire analysis in Section \ref{sec:discussion}, and we provide closing arguments and conclusions in Section \ref{sec:conclusion}.

\section{Discovery and follow-up} \label{sec:discovery}
\begin{figure}
   \centering
   \includegraphics[width=6cm]{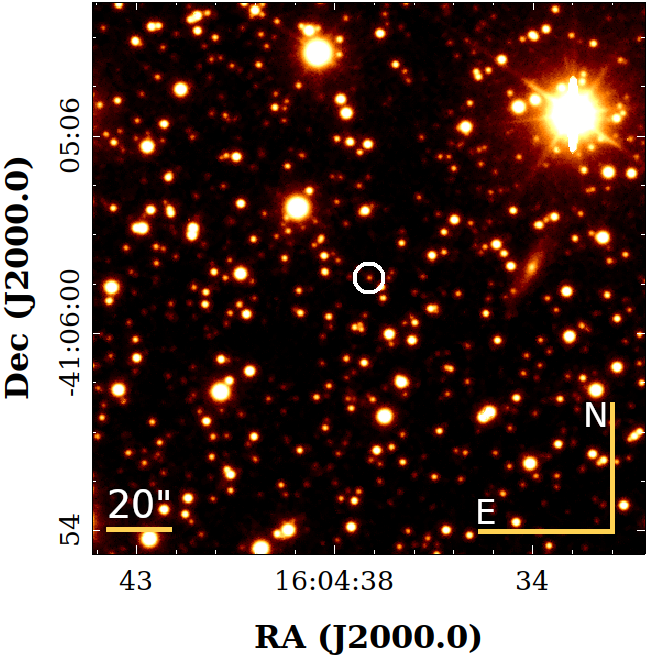}
   \caption{Location of Gaia18cbf and its neighbourhood during the event. Chart is a stack of two images obtained with 2~m LCO telescope located in Siding Springs Observatory in October~2019 and July~2020. White circle marks the target.}
   \label{fig:fchart}
   \end{figure}

Gaia18cbf (AT2018etg according to the IAU transient name server was detected by the Gaia Science Alerts pipeline using the data collected on August 6 2018 (HJD' = HJD - 2450000.0 = 8336.62). The event was located at equatorial coordinates $RA$= 16:04:38.86, $\delta$=-41:06:17.39 and Galactic coordinates $l = 337^\circ\hskip-2pt .594$, $b = 8^\circ\hskip-2pt .413$.
We present a finding chart with the location of this event in Figure \ref{fig:fchart}.
Two days later, on August 8, it was posted on GSA the website\footnote{\href{http://gsaweb.ast.cam.ac.uk/alerts/alert/Gaia18cbf/}{http://gsaweb.ast.cam.ac.uk/alerts/alert/Gaia18cbf/}}. 

Gaia Data Release 2~(GDR2) \citep{GaiaDR2} and Gaia Early Data Release 3~(GEDR3) \citep{2020EDR3, 2020EDR3Astrometry} contain an entry for this source under Gaia Source ID = 5995134605563805056.
We present values of full five-parameter astrometric solutions along with estimated distances from \cite{BailerJones} and \cite{2021Bailer-JonesEDR3} for this object in Table \ref{tab:gdrsVals}.
 
It is important to note that in GDR2 the parallax is negative, while the proper motions were measured with significant uncertainties. The precision of proper motions has improved in GEDR3; however, the parallax value is still negative.
Estimated distances are not reliable for negative parallaxes and very faint sources.
In case of microlensing, if the lens is luminous enough, the parallax measurement may be affected \citep{1999Jeong}.
The lens can dominate over the source's light, and the observed parallax and proper motion can be measured for the mean source and lens motion.
This means that parameters presented in GDR2 and GEDR3 should be taken into account only after we have found a best fitting model for this event and after we have ruled out the case of a luminous lens.

\begin{table}
\caption{\label{tab:gdrsVals}Gaia astrometric parameters for the source star in Gaia18cbf.}.
     \centering
        \begin{tabular}{c c c}
        \hline
        \noalign{\smallskip}
             Parameter &  GDR2 & GEDR3 \\
             \noalign{\smallskip}
        \hline
        \hline
        \noalign{\smallskip}
             $\varpi$ [mas] & $-1.11\pm0.70$ &  $-0.36\pm0.59$ \\
             $\mu_{\alpha}$ [$\mathrm{mas} \, \mathrm{yr}^{-1}$] & $-0.68\pm1.96$ & $-1.83\pm0.68$ \\
             $\mu_{\delta}$ [$\mathrm{mas}\, \mathrm{yr}^{-1}$] & $-0.76\pm1.16$ & $-1.82\pm0.46$ \\
        \noalign{\smallskip}
        \hline
        \noalign{\smallskip}
         \multicolumn{3}{c}{Bailer-Jones et al. distances} \\
        \noalign{\smallskip}
        \hline
        \noalign{\smallskip}
             $r_{\rm est}$ [kpc] &  $4.4^{+3.2}_{-2.9}$ & --\\
             $r_{\rm geo, est}$ [kpc] &  -- & $5.4^{+2.2}_{-1.9}$ \\
             $r_{\rm photgeo, est}$ [kpc] &  -- & $7.8^{+2.0}_{-1.4}$ \\
        \noalign{\smallskip}
        \hline
        \end{tabular}
\tablefoot{Parallax $\varpi$ and proper motions $\mu_\alpha, \mu_\delta$ come from \textit{Gaia} Data Release~2 \citep{GaiaDR2} and \textit{Gaia} Early Data Release~3 \citep{2020EDR3}. The distance estimates come from \cite{BailerJones} and \cite{2021Bailer-JonesEDR3}.}
\end{table}

\subsection{Gaia photometry} \label{sec:phot}

Gaia's photometric measurement is based on the mean of brightness from nine AFs and it is gathered in a wide $G$-band \citep{2010JordiGaiaPhot}. 
On average, Gaia observes a source once every 30 days. 
This event was discovered before the time of peak brightness. 
However, due to its unusual shape and low luminosity, it was first dismissed as a non-microlensing event. 
Only after the new Gaia data points collected in  February 2019, from around the peak in brightness, did the event become a candidate microlensing event. 
As of October 15 2021, Gaia has collected 76 measurements for Gaia18cbf and the event has almost 
returned to its baseline.
Gaia Science Alerts do not provide errors for published events, but the nominal error should vary from 0.02~mag for $G=17.7$~mag at peak to 0.1~mag for $G=20.3$~mag at baseline \citep{GaiaDR2}.

We applied the following formula to provide estimates of error bars for Gaia measurements, since the light curve published by GSA provides none:
\begin{equation*}
  \log_{10}(\mathrm{err}(G)) = 
  \begin{cases}
    0.20\cdot13.5~\text{mag} -5.20~\text{mag}, G \leq 13.5~\text{mag} \\
   0.20\cdot G -5.20~\text{mag}, 13.5~\text{mag} < G < 17.0~\text{mag}\\
   0.26\cdot G -6.26~\text{mag}, G \geq 17.0~\text{mag}. \\
  \end{cases}
\end{equation*}
We present the Gaia data used for modelling in Table \ref{tab:photGaia}.
\subsection{Photometric follow-up}
We require consistent follow-up in order to obtain proper values for a best fitting microlensing model, especially blending parameters.
This means that the event should be observed not only at the time of peak brightness, but also during the baseline by the same telescope.
Due to the target faintness and location, we chose to use the Las Cumbres Observatory Global Network (LCO) of robotic telescopes \citep{LCONetw}.

The follow-up observation started after the peak using the 2-metre~LCO telescope located at Siding Spring Observatory, Australia ($31^\circ16' 23.88''$~S, $149^\circ4'15.6''$~E, LCO SSO).
The number of data points is presented in Table \ref{tab:datapoints}. 
Follow-up data were uploaded and calibrated using the Black Hole Target Observation Manager (BHTOM)\footnote{\href{https://bh-tom.astrolabs.pl/}{https://bh-tom.astrolabs.pl/}} tool for coordinated observations and processing of photometric time series. BHTOM uses 
Cambridge Photometric Calibration Server \citep{2019CPCS2, 2020CPCS2} and CCDPhot (Miko{\l}ajczyk et al., in prep.) as its main processing pipelines.
In order to model this event, we chose to use the observations made in the i band only. This is because there are significantly more points in the $i$ band compared to the $V$ and $G$ bands.
The $i$-band data are presented in Table \ref{tab:photGaiaCSPSi}.

\begin{table}
\caption{\label{tab:datapoints}Summary of photometric data gathered for Gaia18cbf.}
         \centering
         \begin{tabular}{c c c}
            \hline
            \noalign{\smallskip}
            Facility &  Filter &  Data points\\
            \noalign{\smallskip}
            \hline\hline
            \noalign{\smallskip}
            Gaia & $G$ & 76 \\
            \hline
            \multirow{3}{*}{Las Cumbres SSO} & $V$ & 5 \\
             & $i'$ & 34 \\
             & $r'$ & 2 \\
            \noalign{\smallskip}
            \hline
        \end{tabular}
\end{table}

\begin{figure*}
   \centering
   \includegraphics[width=\textwidth]{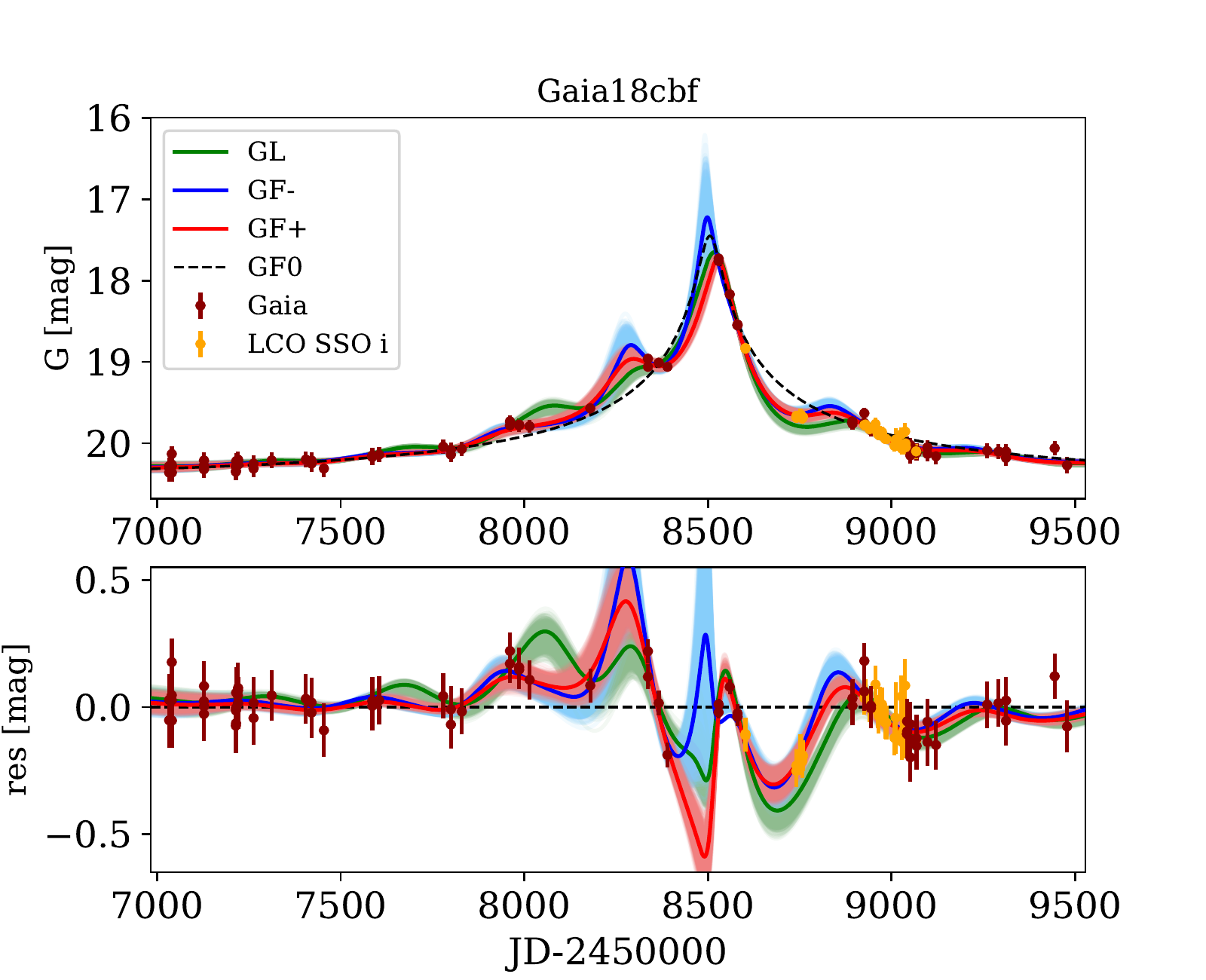}
   \caption{Light curve and best fitting models for Gaia18cbf with follow-up. Maroon dots represent data gathered by Gaia, while yellow dots mark data observed by Las Cumbres Observatory in Siding Spring Observatory (LCO SSO) in the i band. Continuous lines represent different solutions for a point source-point lens model with parallax effect. Best fitting models for \mbox{GF+}, \mbox{GF-} and \mbox{GL} solutions are shown in red, blue, and green, respectively. In lighter colours, we show other solutions that we obtained through a Markov Chain Monte Carlo method. The dashed line represents the point source--point lens model without including a parallax effect, called 'GF0'. On the bottom panel, we show the difference between \mbox{the GF0} model and other models. The data shown in maroon and yellow dots represent the result of subtraction of the magnitude predicted by the GF0 model and the observed magnitude. We use the same colour representation for \mbox{the GF+}, \mbox{GF-,} and \mbox{GL} solutions, as in the top panel.}
   \label{fig:modelAll}
   \end{figure*}

\section{Microlensing model} \label{sec:model}
The standard Paczy{\'n}ski light curve model \citep{1986Paczynski}, which assumes a point source and a point lens, requires three parameters: the time of the peak of brightness $\tnod$, impact parameter $\beta$, which is the lens-source separation (scaled to $\thetae$) at $\tnod$, and the timescale of an event, called Einstein time $\te,$ defined as:
\begin{equation}
    \label{eq:timescale}
    \te = \frac{\thetae}{\mu_{\mathrm{LS}}},
\end{equation}
where $\mu_{\mathrm{LS}}$ is the relative proper motion of source and lens, which is typically unknown.
Then, the time-varying projected lens-source separation, scaled to $\thetae$ is expressed as:
\begin{equation}
\label{eq:positoion}
    \centering
    u(t) = (\tau^2 + \beta^2)^{1/2} ;
    \tau(t) = \frac{t-\tnod}{\te}, \,\,\, 
\end{equation}
which leads to the expression for the magnification of the source star as follows:
\begin{equation}
    \centering
    A(t) = \frac{u(t)^2 + 2}{u(t) \sqrt{u(t)^2 + 4}} .
\end{equation}
Finally, the relation for observed flux as a function of time is:
\begin{equation}\label{eq:ulensFlux}
    \centering
    F(t) = \Fs A(t) + \Fb,
\end{equation}
where $\Fs$ is the unmagnified flux of the source and $\Fb$ is the blend flux. 
The blend flux could be the light coming from the lensing star or, in dense fields, from nearby stars blended in the source image.
If we consider binary lens or binary source models, the blend can also come from the lens' or the source's companion.
Blending can also be a combination of the aforementioned factors.

In this paper, we use different flux parameters: $\Gnod$, $\inod$, $\fsg$ and $\fsi$.
The $\Gnod$ is the brightness in baseline in $G$ filter, which is derived from $\Fs$ and $\Fb$ from equation \ref{eq:ulensFlux} using following relation:
    \begin{equation}\label{eq:blendParamsBase}
        \Gnod = -2.5log_{10}(F_{\mathrm{s},G}+F_{\mathrm{b},G}) + \mathrm{Zero~Point}_G.
    \end{equation}
Similarly, $\inod$ is the brightness in the baseline in the $i$ filter, calculated as $\Gnod$.
The $\fsg$ is the blending parameter in $G$ filter, which is related to $\Fs$ and $\Fb$ by the following relation:
    \begin{equation}\label{eq:blendParamsBlend}
       \fsg = \frac{F_{\mathrm{s},G}}{F_{\mathrm{s},G} + F_{\mathrm{b},G}}. 
    \end{equation}
and $\fsi$ is the blending parameter in the $i$ filter.

\subsection{Annual parallax}
The light curve presented in Figure \ref{fig:modelAll} reveals that the event lasted for over four years.
This means that in modelling this event we have to include the annual parallax effect to take into account Earth's movement around the Sun.

First we have to introduce the parallax vector $\mathbf{\pi_\mathrm{E}}$ \citep{2000ApJ...542..785G}, which has the same direction as the relative proper motion of the lens and source $\mathbf{\mu_\mathrm{LS}}$ and its magnitude is the relative parallax of the lens and source scaled by the Einstein radius:
\begin{equation}
    \overrightarrow{\pie} = \frac{\pi_\mathrm{LS}}{\theta_\mathrm{E}} \frac{\overrightarrow{\mu_\mathrm{LS}}}{\mu_\mathrm{LS}} \,\,,
\end{equation}
where $\pi_\mathrm{LS} = \mathrm{au} (\frac{1}{D_\mathrm{L}} - \frac{1}{D_\mathrm{S}})$ \citep{2004Gould}.
We adopt a geocentric coordinate system and follow equations introduced in \cite{2004Gould}.
We have to add new terms to the microlensing model to take into account that during the year the observer located on Earth changes their position. 
These terms will correspond to changes of the source--lens projected separation and magnification.
Relation \ref{eq:positoion} will then change to
\begin{equation}
    \tau(t) = \frac{t-\tnod}{\te} + \delta t, \,\, \beta(t) = \unod + \delta \beta,
\end{equation}
where
\begin{equation}
    (\delta \tau, \delta \beta) = \pie \overrightarrow{\Delta s} = (\overrightarrow{\pie} \cdot \overrightarrow{\Delta s}, \overrightarrow{\pie} \times \overrightarrow{\Delta s})
\end{equation}
is the displacement vector due to parallax and $\mathbf{\Delta s}$ is the positional offset of the Sun in the geocentric frame \citep{2004Gould}.

\subsection{Modelling and results}

We used a Markov chain~Monte Carlo (MCMC) method to explore the parameter space of the model described above. 
We used the \texttt{emcee} package \citep{2013Emcee} for the MCMC method and the \texttt{MulensModel} package \citep{MulensModel} to generate microlensing models including an annual parallax effect.
We note that \texttt{MulensModel} returns flux parameters in form of $\Fs$ and $\Fb$, as in Equation \ref{eq:ulensFlux}. 
We then calculated $\fsg$ and $\fsi$ using Relations \ref{eq:blendParamsBase} and \ref{eq:blendParamsBlend}.
We assumed $Zero Point_{G}$ and  $Zero Point_{i}$ as 22~mag following the \texttt{MulensModel} package parametrisation.

We modelled the event using two sets of photometric data: Gaia data only and Gaia and i-band follow-up data from LCO SSO. 
We first performed a preliminary exploration of the parameter space to discover possible solutions. 
After visually inspecting the resulting solutions, we found three classes of solutions for Gaia-only data: one with an unusually long Einstein timescale $\te$ (dubbed GL), and two with shorter $\te$ but symmetrical in relation to the impact parameter $\unod$, (dubbed GS+ and GS-). 
For comparison, we also computed a model without including a parallax effect, called 'G0'. 
We present parameters of best fitting models in Table \ref{tab:valSolutionsGaia}.

\begin{figure*}
   \centering
   \includegraphics[width=0.75\textwidth]{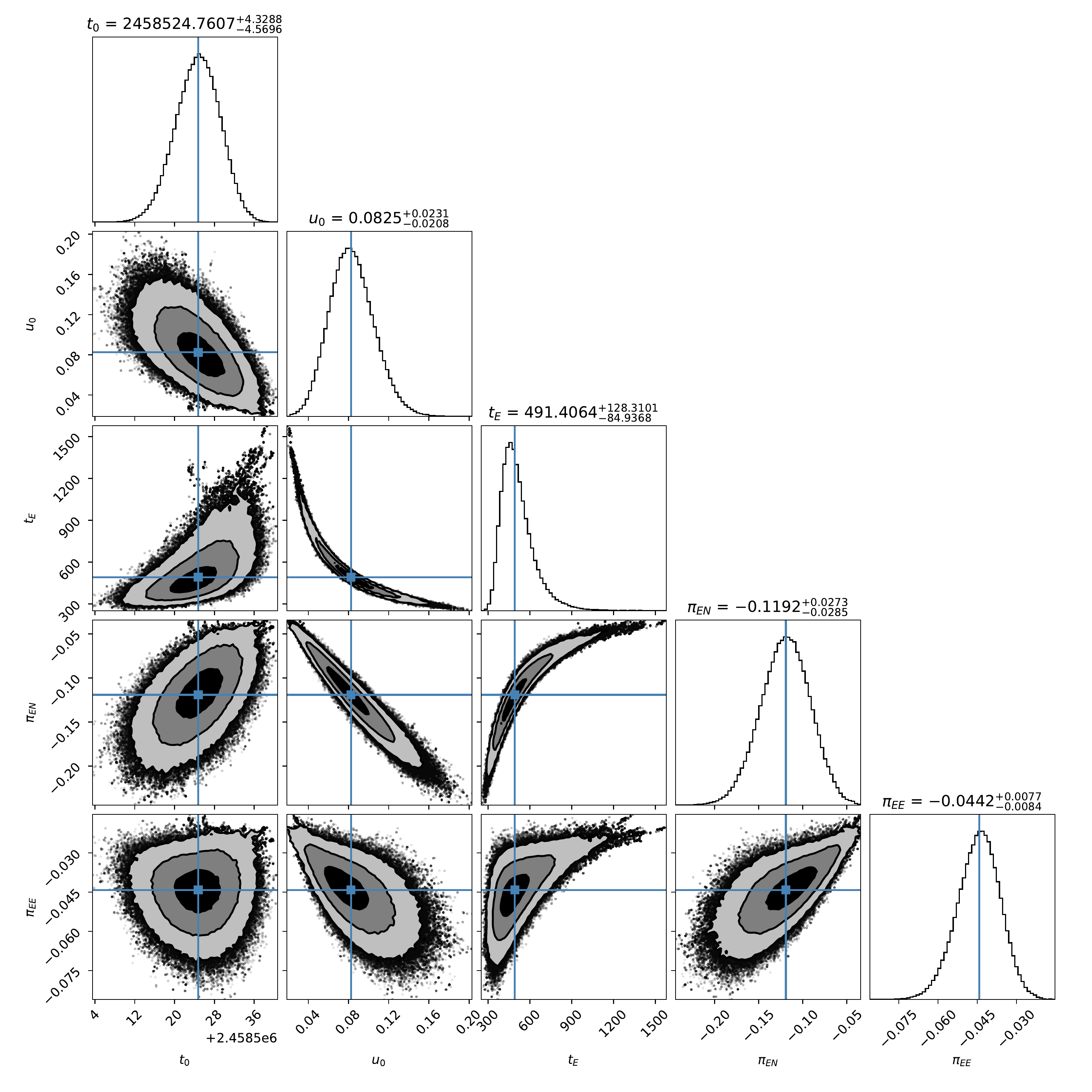}
   \caption{Chi-squared contours plotted as a function of the parameters fitted in the MCMC fit for the best model, \mbox{GF+}. Black, dark grey, and light grey solid colours represent $1 \sigma$, $2 \sigma$, and $3 \sigma$ confidence regions, respectively. Black dots represent solutions outside of the $3 /sigma$ confidence level. Blue lines and squares mark the median solution reported in Table \ref{tab:valSolutionsGaiaFol}. Plot was created using \texttt{corner} python package by \cite{corner}.}
   \label{fig:cornerBest}
   \end{figure*}

When we added the follow-up data to the Gaia data, we noticed that the GL solution disappeared, but two symmetrical, with the regard to impact parameter, $\unod$ solutions remained, called GF+ and GF-. 
We also added a model without including the parallax effect for comparison, called GF0.
We present parameters of the best fitting models for this case in Table \ref{tab:valSolutionsGaiaFol}. The chi-squared contour plots in function of fitted parameters of the best solution, \mbox{GF+}, are presented in Figure \ref{fig:cornerBest}.


The GL solution also has the highest $\chi^2$ value.
The lowest $\chi^2$ value is found for a solution with $\unod>0$ for both datasets used.
When we use both Gaia and i-band follow-up data, the solution with $\unod>0$~(GF+) also has the highest value of blending parameter $f_\mathrm{s}=\Fs/(\Fs+\Fb)$, implying that the observed flux is dominated by the flux from the source and the blended flux is negligible.
This strongly supports the hypothesis that a dark object caused the Gaia18cbf microlensing event.
Other solutions, however, have blending parameter values varying between 0.6 and 0.8, which would suggest that the lens is a dim star rather than a dark remnant of stellar evolution.
We present the Gaia and follow-up data along with the best fitting light curves in Figure \ref{fig:modelAll}.

\begin{table*}
    \centering
    \caption{\label{tab:valSolutionsGaia}Parameter values of best fitting solutions for Gaia18cbf using only Gaia photometric data.}
    \begin{tabular}{c c c c c}
            \hline
            \noalign{\smallskip}
            Parameter & G0 & GL & GS+ & GS- \\
            \noalign{\smallskip}
            \hline
            \hline
            \noalign{\smallskip}
            $t_\mathrm{0, par}-245000.$ [days]  & -- & \multicolumn{3}{c}{$8529.00$} \\
            $\tnod$ [days] & $8505.17^{+2.02}_{-2.01}$ &  $ 8388.56^{+21.89}_{-20.58}$ & $8528.46^{+4.37}_{-4.85}$ & $8516.64^{+4.18}_{-4.82}$\\
            \noalign{\smallskip}
            $\unod$ & $0.0131^{+0.0087}_{-0.0064}$ & $-0.0023^{+0.0211}_{-0.0205}$ & $0.0828^{+0.0253}_{-0.0223}$ & $-0.0549^{+0.0164}_{-0.0189}$ \\
            \noalign{\smallskip}
            $\te$ [days] & $2077.74^{+1975.26}_{-798.44}$ & $2347.96^{+929.41}_{-583.25}$ & $487.26^{+135.58}_{-88.23}$ & $476.53^{+191.52}_{-118.71}$ \\
            \noalign{\smallskip}
            $\pien$ & -- & $0.0458^{+0.0218}_{-0.0226}$ & $-0.1280^{+0.0305}_{-0.0324}$& $-0.1701^{+0.0518}_{-0.0623}$ \\
            \noalign{\smallskip}
            $\piee$ & -- & $-0.12801^{+0.0366}_{-0.0416}$ & $-0.0519^{+0.0103}_{-0.0115}$ & $-0.0332^{+0.0080}_{-0.0081}$\\
            \noalign{\smallskip}
            $\Gnod$ [mag] & $20.44^{+0.07}_{-0.07}$ & $20.34^{+0.04}_{-0.03}$ & $20.33^{+0.04}_{-0.03}$ & $20.34^{+0.04}_{-0.03}$ \\
            \noalign{\smallskip}
            $\fsg$ & $0.196^{+0.110}_{-0.090}$ & $0.666^{+0.221}_{-0.191}$ & $0.888^{+0.279}_{-0.234}$ & $0.647^{+0.279}_{-0.205}$ \\
            \noalign{\smallskip}
            $\chi^2$ & 117.79 & 36.48 & 27.15 & 31.55 \\
            $\frac{\chi^2}{dof}$ & 1.66 & 0.53 & 0.39 & 0.46 \\
            \noalign{\smallskip}
            \hline
    \end{tabular}
    \tablefoot{G0 is the point source--point lens model without including a microlensing parallax effect. GL (Gaia-long), GS+ (Gaia-short-positive $\unod$) and GS- (Gaia-short-negative $\unod$) are the point source--point lens models including a microlensing parallax effect. For G0, we chose the model with $\unod>0$, because the posterior distribution for $\unod$ was symmetrical.}
\end{table*}

\begin{table*}
        \centering
        \caption{\label{tab:valSolutionsGaiaFol}Parameter values of best fitting solutions for Gaia18cbf using Gaia and i-band follow-up photometric data.}
         \begin{tabular}{c c c c}
            \hline
            \noalign{\smallskip}
            Parameter & GF0 & GF+ & GF- \\
            \noalign{\smallskip}
            \hline
            \hline
            \noalign{\smallskip}
            $t_\mathrm{0, par}-245000.$ [days] & -- & \multicolumn{2}{c}{$8529.00$} \\
            $\tnod-2450000.$ [days] & $8505.34^{+2.04}_{-2.00}$ & $8524.76^{+4.33}_{-4.57}$ & $8513.14^{+4.68}_{-3487}$ \\
            \noalign{\smallskip}
            $\unod$ & $0.0133^{+0.0086}_{-0.0065}$ & $0.0825^{+0.0231}_{-0.0208}$ & $-0.0567^{+0.0162}_{-0.0169}$ \\
            \noalign{\smallskip}
            $\te$ [days] & $2058.40^{+1956.82}_{-781.12}$ & $491.41^{+128.31}_{-84.94}$ & $453.74^{+178.69}_{-105.74}$\\
            \noalign{\smallskip}
            $\pien$ & -- &$-0.1192^{+0.0273}_{-0.0285}$ & $-0.1697^{+0.0516}_{-0.0577}$ \\
            \noalign{\smallskip}
            $\piee$ & -- & $-0.0442^{+0.0077}_{-0.0084}$ & $-0.0257^{+0.0054}_{-0.0057}$\\
            \noalign{\smallskip}
            $\Gnod$ [mag] & $20.44^{+0.07}_{-0.07}$ & $20.34^{+0.03}_{-0.03}$ & $20.34^{+0.04}_{-0.03}$\\
            \noalign{\smallskip}
            $\fsg$ & $0.198^{+0.109}_{-0.090}$ & $0.894^{+0.263}_{-0.226}$ & $0.717^{+0.294}_{-0.233}$ \\
            \noalign{\smallskip}
            $\inod$ [mag] & $20.06^{+0.07}_{-0.07}$& $20.01^{+0.06}_{-0.05}$ & $20.00^{+0.07}_{-0.06}$\\
            \noalign{\smallskip}
            $\fsi$ & $0.157^{+0.084}_{-0.072}$ & $0.996^{+0.250}_{-0.219}$ & $0.746^{+0.253}_{-0.208}$\\
            \noalign{\smallskip}
            $\chi^2$ & 143.98 & 40.87 & 46.69 \\
            $\frac{\chi^2}{dof}$ & 1.40 & 0.40 & 0.46 \\
            \noalign{\smallskip}
            \hline
        \end{tabular}
        \tablefoot{GF0 is the point source--point lens model without including a microlensing parallax effect. GF+ and GF- are the point source--point lens models with including a microlensing parallax effect. For GF0, we chose the model with $\unod>0$, because the posterior distribution for $\unod$  was symmetrical.}
\end{table*}

\section{Source star} \label{sec:source_star}

To determine the physical properties of the lensing event, a distance estimation for the source star is essential.
The most obvious place to find the distance of a star, in particular in case of an event discovered by Gaia, are the Gaia data releases. 
In this case we could not, however, use distances from GDR2 or GEDR3, because the parallax $\varpi$ is negative and thus unreliable.
Instead, we used archival photometric data and applied a spectro-photometric method to obtain the distance of the source star.
This method assumes that the star is single and non-variable.
We also assumed that the extinction is proportional to the colour excess.
Additionally, we assumed that the light observed at all stages of the microlensing event is emitted by a single object, the background (source) star. 
The microlensing model can provide information on the amount of blended light (from lens or unrelated stars within the seeing disc), and, as shown in Section \ref{sec:model}, different models indicate different values of blending.
However, only in extreme cases can the blend dominate over the source; hence, in this study of the source we assumed the light at the baseline is dominated by the source star. 
In our case, the blending parameter $f_s$ does not fall below 0.6, so we can assume that the majority of the light during the event comes from the source.


In this work, we used the VISTA Hemisphere Survey~(VHS) catalogue DR4.1 \citep{2013McMahon} and the TESS Input Catalog--v8.0~(TIC-8) \citep{2019Stassun}.
The VISTA catalogue provides J and Ks magnitude values, as well as extinction, which allowed us to determine that the intrinsic colour of the source is star equal to $(J-K)_0=0.131$~mag.
According to \cite{2013PecautMamajek} (hereafter PM), this colour most closely corresponds to a F0V-type star with an effective temperature of $T_\mathrm{eff}=7220$~K and $M_V=2.51$~mag.
For comparison, we empirically determined the intrinsic colours $(J-H)_0$ and $(H-Ks)_0$ using the 2MASS system \citep{2009StraizysLazauskaite}, which provided us a F0V star value equal to $(J-K)_0=0.135$~mag.
Using the $V$ magnitude from the TESS catalogue, we computed $V_0$ taking into account the reddening  $E(B-V)=0.482$~mag. 
We adopted the coefficient $R_V= 3.15$, which corresponds to the normal extinction law equal to $A_V=1.534$~mag.
This gave us $V_0=18.955$~mag and $(V-K_s)_0=0.785$~mag.
These values are colour indices in the middle of F0V and F1V spectral class. 
In our case, the uncertainties on magnitudes are $e_V=0.062$~mag, $e_J=0.081$~mag and $e_{K_s}=0.279$~mag.
However, because the error for $K_s$ is larger than the other two, we decided to use the $(V-J)_0$ colour for comparison.
We calculated it using following relation:
\begin{equation}
(V-J)=(V-K_s)-(J-K_s) 
\end{equation}
and \citetalias{2013PecautMamajek} data.
Our obtained colour value of $(V-J)_0 = 0.654$~mag again corresponds to a mean value between F0V~(0.589mag) and F1V~(0.662mag) stars. 
According to the three obtained colour index parameters, we assume that our source star spectral type is either F0V or F1V, which gives us absolute magnitudes of $M_V=2.51$~mag or $M_V=2.79$~mag, respectively.
For further distance determination, we used an average value of $M_V=2.65\pm0.50$~mag, including a conservative error due to unknown metallicity of the star.
We determined the distance $D_\mathrm{S}$ for the source star using the following relation:
\begin{equation}
    \log_{10}(D_\mathrm{S}) = \frac{(V_0 - M_V + 5)}{5}.
\end{equation}
We deduced thereby that the distance to the source star is 18.2~kpc.

The errors of the observed magnitudes, colour excesses, and extinction usually follow a Gaussian distribution, but taking into account non-linearity of the transformation and asymmetry of the distance ($\log_{10}(D_\mathrm{S})$), we conclude the distance to the star is between 14.5~kpc and 23~kpc, which is in contrast with \cite{BailerJones} and \cite{2021Bailer-JonesEDR3} ($1.\,\mathrm{kpc} < D_\mathrm{S} < 7.6)$~kpc and $6.4\,\mathrm{kpc}<D_\mathrm{S}<9.8$~kpc, respectively).
We note that the exact value of the distance has, in practice, a small effect on the final models because $\pi_s \ll \theta_E \pi_E$.
However, it is important to differentiate between nearby and faraway sources. 

The uncertainties in distance estimation using this method are large.
However, due to the dimness of the sources, we could not obtain a spectrum that would result in a more precise distance estimation.

We also derived the value of the interstellar extinction in the $G$ band, $A_G$, towards this source.
Since GDR2 does not provide any measurement of $A_G$ for Gaia18cbf, we instead calculated the average $A_G$ of all sources within 1~arcsecond of the source (8 stars in total), finding $A_G = 1.06$~mag.
We compared this value with extinction provided by \cite{2011Schlafly} for SDSS bands: $A_{\mathrm{SDSS}, g}=1.810$~mag,  $A_{\mathrm{SDSS}, r}=1.252$~mag, and $A_{\mathrm{SDSS}, i}=0.931$~mag. 
Since the Gaia $G$-filter bandpass is wide \citep{2010JordiGaiaPhot, 2018EvansPhoto, 2021RielloPhoto}, spanning from 300nm to 1000nm, we conclude that the calculated value of extinction is correct.

\section{Lens light analysis} \label{sec:lens}

In order to determine the nature of the lens in this event, we followed the procedure used in \cite{2016WyrzykBH} and \cite{MW} (hereafter MW). In short, the microlensing parameters in the MCMC samples from the light-curve modelling were combined with the priors on missing parameters: namely, source distance, source proper motion, and lens proper motion. 
Since the blending parameter found in our main parallax solutions (\mbox{GF+} and \mbox{GF-}) was above 0.7 in both cases, we adopted the source proper motions measured by Gaia EDR3. For the source distance, we did not use the distance provided in \cite{2021Bailer-JonesEDR3} as it was derived based on strongly negative geometric parallax and biased towards closer distances. 
Instead, for each iteration of the algorithm, we drew the source distance from the flat distribution covering the interval obtained through the analysis of archival photometric data in Sect. \ref{sec:source_star} as well as the extinction value in the G band derived there.

Then we drew a random value of lens-source relative proper motion $\mu_\mathrm{LS}$ from a flat distribution of values between 0 and 30~$\mathrm{mas} \, \mathrm{year}^{-1}$ \citepalias{MW}. 
The combination of $\pi_\mathrm{E}$, $t_\mathrm{E,}$ and $\mu_\mathrm{LS}$ allowed us, for each MCMC sample, to compute the mass of the lens and, in combination with the source distance, its distance. 
This was in turn used to calculate the lens brightness if the lens was a main-sequence (MS) star. 
For our mass-luminosity relation, we used empirical data from \citetalias{2013PecautMamajek} provided on Eric Mamajek's webpage 
\footnote{\href{http://www.pas.rochester.edu/~emamajek}{http://www.pas.rochester.edu/\~{}emamajek}}.
We then compared the luminosity of the lens for a given mass with the blended light derived for a given MCMC sample.
We assumed an extinction value for the source as calculated in Section \ref{sec:source_star}.
For the lens, we adopted a more sophisticated approach. 
We extracted reddening $E(B-V)$ values in the direction of the event using online 3D maps of interstellar medium provided by \cite{Reddening}, \cite{Reddening2}, and \cite{Reddening3}\footnote{\href{https://stilism.obspm.fr/}{https://stilism.obspm.fr/}}.
Using this information, we assume that up to 300~pc the reddening $E(B-V)=0$~mag and than it changes according with a fitted linear function.
Based on that, we calculated $A_G$ using relation presented in \cite{redToExtinct}; however, we did not allow the extinction to be larger than $A_G=1.06$~mag.
For all the pairs $\ml$-$\dl$ and corresponding lens light - blend light, we also computed a weight using set of priors from \cite{2011Skowron} and \citetalias{MW}. 
These priors require $\te$ and $\pie$ calculated in the heliocentric frame. 
We used formulas from Appendix~A of \cite{2011Skowron} to find the correct values of these parameters. 
For the mass function slope, we assumed a power index of $-1.75$, as suggested in \citetalias{MW}.

\begin{figure}
   \centering
   \includegraphics[width=9cm]{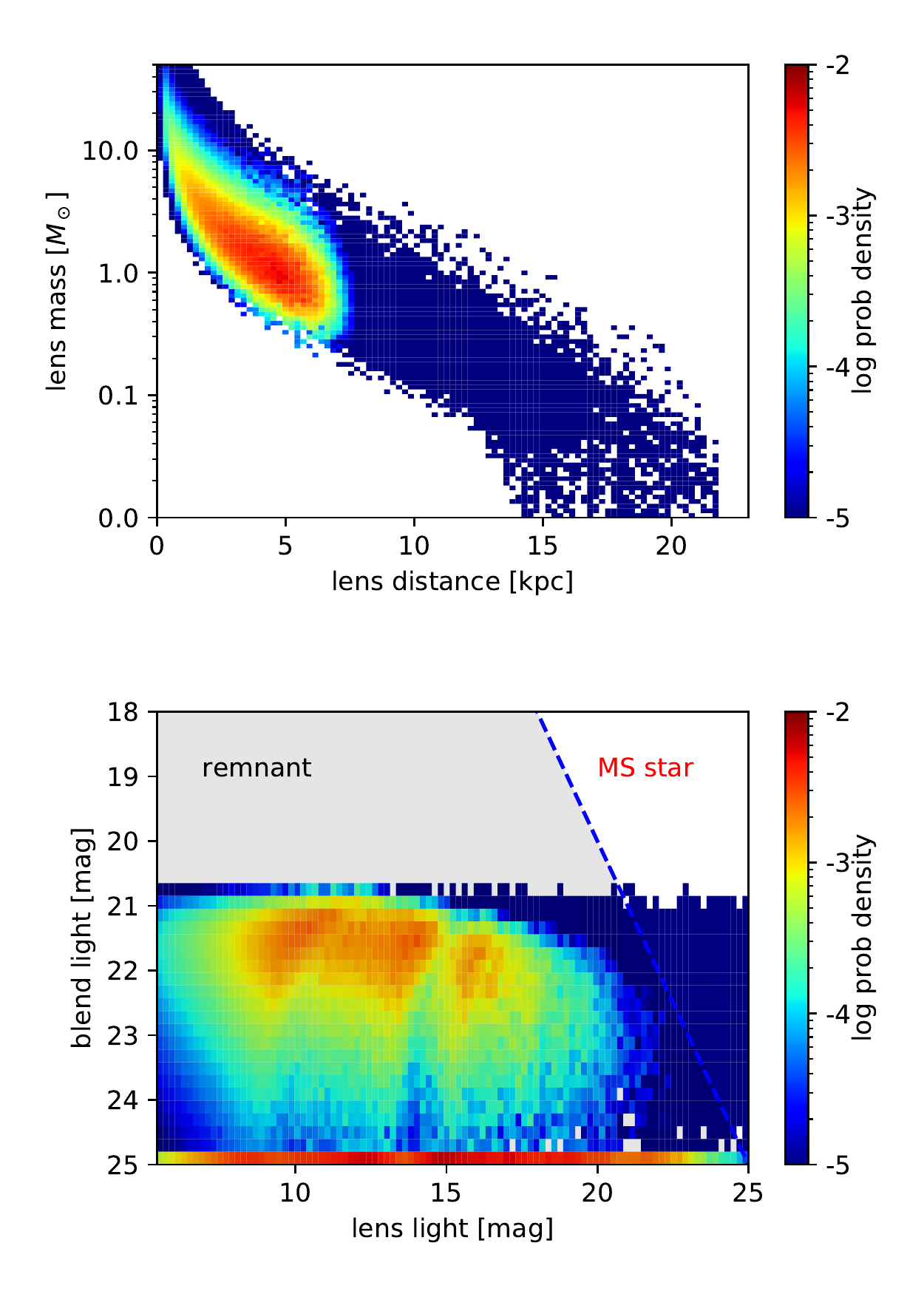}
   \caption{Posterior distribution of the lens mass, distance, lens light if it was a MS star and allowed blend light obtained using method outlined in Section \ref{sec:lens}.
   Top: Lens mass versus lens distance estimated from MCMC samples for Gaia18cbf for the best fitting model using the Gaia and follow-up data with $\unod>0$ and including a microlensing parallax effect, called GF+. Bottom: Blended light magnitude from the microlensing model GF+ versus the light of the lens expected if the lens was a MS star. The dashed line separates dark lens or remnant solutions from MS star solutions. The colours are the probability density shown on a logarithmic scale.}
    \label{fig:massDistanceGF+}
   \end{figure}

\begin{figure}
   \centering
   \includegraphics[width=9cm]{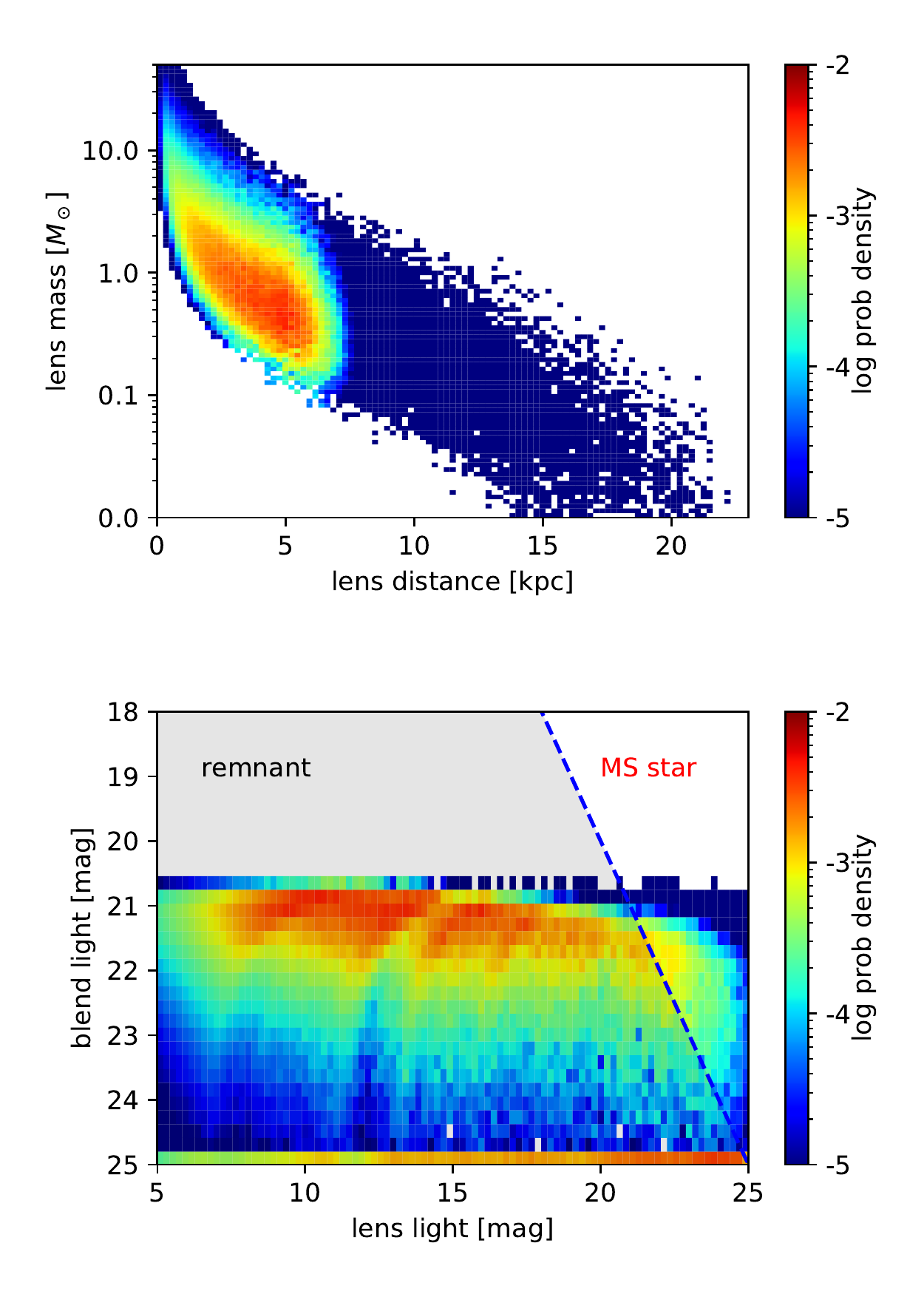}
   \caption{Posterior distribution of the lens mass, distance, lens light if it was a MS star and allowed blend light obtained using method outlined in Section \ref{sec:lens}.
   Top: Lens mass versus lens distance estimated from MCMC samples for Gaia18cbf for the best fitting model using the Gaia and follow-up data with $\unod<0$ and including a microlensing parallax effect, called GF-. Bottom: Blended light magnitude from the microlensing model versus the light of the lens expected if the lens was a MS star. The dashed line separates dark lens or remnant solutions from MS star solutions. The colours are the probability density shown on a logarithmic scale.}
   \label{fig:massDistanceGF-}
   \end{figure}

The results of the lens-light analysis for GF+ and GF- models are shown in Figures \ref{fig:massDistanceGF+} and \ref{fig:massDistanceGF-}, respectively, with the top panels showing the probability density for mass and distance of the lens. 
Table \ref{tab:lensParams} gathers the most likely values for mass and distance of the lens in both solutions, as well as the value of the blend magnitude. 
For GF+ the most likely mass is $\ml = 2.65^{+5.09}_{-1.48} M_\odot$ with the distance of $\dl = 2.84^{+1.94}_{-1.67}$~kpc.
For GF-, the most likely mass is $\ml = 1.71^{+3.82}_{-1.06} M_\odot$ with the distance of $\dl = 2.66^{+1.97}_{-1.56}$~kpc.
The more massive GF+ solution also has a larger value of the blending parameter $f_s$ (hence less light remains for the blend/lens), when compared to the GF- solution. 
The median distance, however, is similar for both solutions and depends mostly on the distance to the source. 

The bottom panels of Figs. \ref{fig:massDistanceGF+} and \ref{fig:massDistanceGF-} display the comparison between the blend light (as measured in the microlensing model) with the light expected for a MS star if the lens was fully responsible for the observed blending light. The dashed line divides the parameter space for a case where the MS scenario is justified given the blending, and where the MS scenario does not explain the mass for the given blending, and hence the remnant scenario is preferred. Given the faint baseline magnitude and relatively high value of the blending parameter, there is not much light in the light curve that could be attributed to the possible lens. 

\begin{table}
        \centering
        \caption{
        \label{tab:lensParams}Lens masses $\ml$ and distances $\dl$ for best fitting models for Gaia18cbf.}
        \begin{tabular}{c c c}
            \hline
            \noalign{\smallskip}
            Parameter & GF+ & GF- \\
            \noalign{\smallskip}
            \hline
            \hline
            \noalign{\smallskip}
             $\Gbl$~[mag] & $21.31^{+\inf}_{-0.47}$ & $20.89^{+\inf}_{-0.23}$ \\
             \noalign{\smallskip}
             \hline
             \noalign{\smallskip}
\multicolumn{3}{c}{Mass function $\propto M^{-1.75}$, $14.5~\mathrm{kpc} < \ds < 23.0~\mathrm{kpc}$} \\
             \noalign{\smallskip}
            \hline
            \noalign{\smallskip}
            $\ml \, [M_\odot]$ & $2.65^{+5.09}_{-1.48}$ & $1.71^{+3.78}_{-1.06}$ \\
            \noalign{\smallskip}
            $\dl$~[kpc] & $2.84^{+1.94}_{-1.67}$ & $2.66^{+1.97}_{-1.56}$\\
            \noalign{\smallskip}
            $G_\mathrm{blend, MS}$~[mag] &  13.94 (B9.5V) &   15.33 (F0V) \\
            \hline
        \end{tabular}
\tablefoot{Here, we assumed the distance to source star found in Section \ref{sec:source_star}. $\Gbl$ is the limit for the brightness of the lens and $G_\mathrm{blend, MS}$ is brightness of the lens if it was a MS star of mass similar to $\ml$ located at the median distance $\dl$. We assumed that extinction changes with the distance (see Section \ref{sec:lens}). The absolute magnitude in the $G$ band was taken from \citetalias{2013PecautMamajek}, and we provide the spectral type of the object used for our calculations in brackets.}
\end{table}

\begin{table}
        \centering
        \caption{
        \label{tab:lensScenarios}Lens masses $\ml$ and distances $\dl$ for best fitting solutions assuming different distances to the source or lens mass distribution.}
        \begin{tabular}{c c c}
            \hline
            \noalign{\smallskip}
            Parameter & GF+ & GF- \\
            \noalign{\smallskip}
            \hline
            \hline
            \noalign{\smallskip}
            $\Gbl$~[mag] & $21.31^{+\inf}_{-0.47}$ & $20.89^{+\inf}_{-0.23}$ \\
             \noalign{\smallskip}
             \hline
            \noalign{\smallskip}
            \hline
            \noalign{\smallskip}
            \multicolumn{3}{c}{Mass function $\propto M^{-1.00}$, $14.5~\mathrm{kpc} < \ds < 23.0~\mathrm{kpc}$} \\
            \noalign{\smallskip}
            \hline
            \noalign{\smallskip}
            $\ml \, [M_\odot]$ & $6.17^{+14.79}_{-4.09}$ &  $4.83^{+13.30}_{-3.41}$ \\
            \noalign{\smallskip}
            $\dl$~[kpc] & $1.55^{+2.08}_{-1.02}$ &  $1.45^{+2.00}_{-0.97}$ \\
            \noalign{\smallskip}
            $G_\mathrm{MS}$~[mag] &  <10.76 (B2.5V) &  10.95 (B5V) \\
            \noalign{\smallskip}
            \hline
            \noalign{\smallskip}
            \multicolumn{3}{c}{Mass function $\propto M^{-3.00}$, $14.5~\mathrm{kpc} < \ds < 23.0~\mathrm{kpc}$} \\
            \noalign{\smallskip}
            \hline
            \noalign{\smallskip}
            $\ml \, [M_\odot]$ & $1.23^{+1.32}_{-0.53}$ & $0.62^{+0.80}_{-0.30}$ \\
            \noalign{\smallskip}
            $\dl$~[kpc] & $4.34^{+1.27}_{-1.74}$ & $4.14^{+1.35}_{-1.67}$ \\
            \noalign{\smallskip}
            $G_\mathrm{MS}$~[mag] & 17.81 (F6V) & 21.89 (K8V) \\
            \noalign{\smallskip}
            \hline
            \noalign{\smallskip}
            \multicolumn{3}{c}{Mass function $\propto M^{-1.75}$, $6.4~\mathrm{kpc} < \ds < 9.8~\mathrm{kpc}$} (B-J distance) \\
            \noalign{\smallskip}
            \hline
            \noalign{\smallskip}
             $\ml \, [M_\odot]$ &  $3.78^{+7.78}_{-2.41}$ & $2.38^{+5.73}_{-1.60}$ \\
             \noalign{\smallskip}
            $\dl$~[kpc] & $1.76^{+1.51}_{-1.01}$ & $1.67^{+1.47}_{-0.97}$ \\
            \noalign{\smallskip}
            $G_\mathrm{MS}$~[mag] & -0.39 (B7V) & 13.13 (A0V) \\
            \noalign{\smallskip}
            \hline
        \end{tabular}
        \tablefoot{$\Gbl$ is the limit for the brightness of the lens, $G_\mathrm{MS}$ is the brightness of the lens if it was a MS star of mass similar to $\ml$ located at the median distance $\dl$. We assumed that extinction changes with the distance (see Section \ref{sec:lens}). The absolute magnitude in the $G$ band was taken from \citetalias{2013PecautMamajek}, we provide the spectral type of the object used for calculations in brackets.}
\end{table}

\section{Discussion} \label{sec:discussion}



The event Gaia18cbf clearly exhibited a very long-lasting increase in brightness, which went on for almost 2000 days. 
The most plausible explanation for the shape of the light curve is microlensing by a single lens, affected by the microlensing parallax effect due to the Earth's orbital motion.
The standard microlensing model (shown with a dashed line in Fig.\ref{fig:modelAll}) is not capable of reproducing the data, however, and the follow-up data filling the data gap at around JD=2458750 help to better constrain the parallax model.
Nevertheless, the discrepancy between \mbox{GF+} and \mbox{GF-} models would probably have been resolved if the event had been followed-up nearer to the time of maximum brightness. 
Unfortunately, the part of the sky in which this event occurred is about 8 degrees away from the Galactic plane and was not observed by OGLE or any other surveys at that time. 

As already mentioned in Section \ref{sec:model}, the best results in terms of $\chi^2$ were given by the \mbox{GF+} model.
Both solutions have, however, large $\te$, making this event second longest known microlensing event, with \cite{2002MaoLongUlens} being the first.

In our model, we did not include the effect of space parallax due to the difference in Gaia's location at L2 and ground-based observatories. 
In order to see the shift between light curves observed with Earth-based and space-based observatories, we would have to have a high-amplification event sampled densely at the peak or a caustic crossing. Moreover, the parallax signal measured is small, and hence the impact of 1\% of observer position difference would not be detectable. 

The probability density depends on the assumed mass function slope, and in our mass-light analysis we assumed that the mass function for the lens follows the slope of $\propto M^{-1.75}$, as suggested in \citetalias{MW}. However, the actual value of this slope is poorly known (e.g. \cite{BastianIMF}), and if we assume no prior on the mass (i.e. power of $-1$), the lens mass moves to higher values of  $\ml=6.17^{+14.79}_{-4.09}~M_\odot$ and $\ml=4.83^{+13.30}_{-3.41}$ for GF+ and \mbox{GF-}, respectively. The lens distance changes to smaller values of $\dl=1.55^{+2.08}_{-1.02}~\mathrm{kpc}$ and $\dl=1.45^{+2.00}_{-0.97}~\mathrm{kpc}$. 
Conversely, an extreme slope of $-3$ discourages larger masses and for the Gaia18cbf solutions GF+ and \mbox{GF-} yields $\ml=1.23^{+1.32}_{-0.53}$ and $\ml=0.62^{+0.80}_{-0.30}$. 
For the \mbox{GF-} model, the lens distance increases to  $\dl=4.14^{+1.35}_{-1.67}~\mathrm{kpc}$, which further discourages a massive remnant possibility, as a M2.5V-type MS star at that distance would have 26.6~mag and could hide within the allowed blending of $\Gbl=20.89_{-0.23}^{+\inf}~\mathrm{mag}$.
However, for the \mbox{GF}+ solution, which we find more favourable, the MS scenario with $\dl=4.34^{+1.27}_{-1.74}~\mathrm{kpc}$ still cannot explain the observed blended light of $\Gbl=21.31_{-0.47}^{+\inf}~\mathrm{mag}$.
High-resolution imaging with sensitivity to stars down to 26~mag would allow us to resolve this issue. 

As we mentioned in Section \ref{sec:source_star}, Gaia parallax~$\varpi$ cannot be used for distance estimation in microlensing events.
However, if we wanted to use the values from \cite{2021Bailer-JonesEDR3}, the lens would become more massive and slightly further away than in our initial assumptions. 
We present obtained values of lens mass $\ml$ and lens distance $\dl$ in Table \ref{tab:lensScenarios}.

Figures \ref{fig:massDistanceGF+} and \ref{fig:massDistanceGF-} suggest that the lens is a dark remnant. 
Cases where the lens is a MS star have lower probability compared to the remnant solutions to cause the Gaia18cbf event.
However, we can still test the assumption that the event was caused by a low-mass MS star located further away.
If we assume that the extinction is $A_G=1.06$~mag and the lens is an M5V MS star with an absolute magnitude of $M_G=12.04$~mag \citepalias{2013PecautMamajek}, then the the minimal values of observable brightness of the lens, $\Gbl$, presented in Section \ref{sec:lens}, would translate to distances of $\dl\approx440$~pc for \mbox{GF+} and  $D_\mathrm{L}\approx360$~pc for \mbox{GF-}.
However, such a lens would produce a larger microlensing parallax of $\pie = 1.33$ for \mbox{GF+} and $\pie = 1.46$ for \mbox{GF-} if we assume that the source is 18.2~kpc away.
From the analysis of the light curve, we obtained $\pie = 0.1274^{+0.0288}_{-0.0278}$  and $\pie = 0.1719^{+0.0572}_{-0.0513,}$  respectively.
We can move the lens away; however, $\pie$ will never be smaller than 0.4.
To reach the required parallax value, the lens has to weigh more than $2 M_\odot$.
If we wanted to explain our event with an A- or B-type MS star as the lens, it would have to be at least 30~kpc away to reach maximal brightness allowed by $\Gbl$ of this event.
This means that this event was most likely not caused by a MS star.

This leads us to a conclusion that we have detected a dark remnant candidate.
Based on the masses we obtained, the lens could be a candidate white dwarf, neutron star, or a black hole.
If the hypothesis of the dark remnant is correct, the mass of the object suggests a massive neutron star or a mass-gap object \citep{2010OzelBHmass, 2011FarrBHmass}.
This could be another case that supports the hypothesis that there is in fact no mass gap \citep{2012KreidbergNomassgap} along with recently observed mass-gap objects \citep{2019ThompsonBH, 2020AbbotGWmassgap, 2021UnicornBH}. 

We assumed that the relative proper motion of lens and source can reach up to 30~$\text{mas}\,\text{yr}^{-1}$.
The maximum value can be reached if source and lens are moving in exactly opposite directions and we are dealing with a fast moving lens.
Natal kicks of neutron stars can typically reach up to 420~$\text{km}\,\text{s}^{-1}$ \citep{2005HobbsNS}, which translates to proper motion of $\mu=(40-80)\,\text{mas}\,\text{yr}^{-1}$ for an object located at 1--2~kpc.
If the relative proper motion between the lens and source was this large, the Einstein radius would have to be enormous ($\thetae > 40$~mas) in order to reach an Einstein timescale of over 400~days.
This would result in a more massive lens, however.
In our case, the Einstein radius for the \mbox{GF+} solution is $\thetae= 2.53^{+4.35}_{-1.27}$~mas and for the \mbox{GF-} is $\thetae = 2.11^{+3.73}_{-1.12}$~mas.
This corresponds to relative proper motions of $\mu_\mathrm{LS} = 1.84^{+2.08}_{-0.72}\text{mas}\,\text{yr}^{-1}$ and $\mu_\mathrm{LS} = 1.60^{+1.51}_{-0.61}\text{mas}\,\text{yr}^{-1,}$ respectively.

We cannot exclude the possibility that this event was caused by a binary lens; however, the data are too infrequent to obtain a good model.
If the lens is a binary system, the caustic crossing could happen during the large gap in data before the peak, which lasts 139.31~days.
Assuming that the distance to the source is $\ds = 18.2$~kpc and using the lens and microlensing event parameters that we obtained, the separation of the binary system would have to be $a_{GF+}\leq2.72$~au for \mbox{a GF+} solution and $a_{GF-}\leq1.33$~au for \mbox{a GF-} solution.
If the source is closer, the maximum separation would get smaller.
This means that instead of a single lens, we could be dealing with a binary system consisting of two objects; for example, two neutron stars or two white dwarfs.

If we would like to confirm that the event was in fact caused by a dark remnant, we would have to wait ten  years and use the Extremely Large Telescope to take high angular-resolution follow-up observations of this field to try and identify the source and lens as separate objects. 
If the lens is a neutron star, it may be emitting radio signals that are too weak to be detected by modern-day telescopes.
We could try observing this area with more sensitive radio telescopes in the future.
However, it is worth noting that only a small fraction of neutron stars can be observed as pulsars due to the fact that one of the magnetic poles has be aligned with our line of sight.
If the lens consists of two neutron stars, the system might emit gravitational waves that could be detectable by the upcoming ESA mission, Laser Interferometer Space Antenna \citep{2017LISA}.
Another option would be to wait for the final Gaia data release. 
Based on \cite{2018MNRAS.476.2013R}, we expect that accuracy in the along scan direction (AL) should be between 0.3~mas for the brightest points and 3~mas for baseline magnitude. 
The largest astrometric displacement of the centroid of light due to microlensing occurs for $u(t) = \sqrt{2}$ and is equal to $\delta_\mathrm{max} \approx 0.354 \thetae$.
This gives us maximal astrometric shift spanning between 0.45~mas and 2.44~mas for the \mbox{GF+} and 0.35~mas and 2.07~mas for the \mbox{GF-}.  
This means that astrometric microlensing might be marginally measurable by Gaia.


\section{Conclusion} \label{sec:conclusion}
We presented an analysis of the Gaia18cbf microlensing event, discovered by the Gaia Science Alerts programme, operating within the Gaia space mission. 
The event exhibited one of the longest microlensing timescales ever seen as well as a significant annual microlensing parallax effect. 
The event has two best fitting solutions with a negative and positive impact parameter, $\unod$, which we were not able to resolve due to insufficient follow-up data.
The parameters of these solutions are presented in Table \ref{tab:valSolutionsGaiaFol}.

Using a Galactic-density and kinematics model, we estimated the probability density of the lens mass and distance.
We present these densities in Figures \ref{fig:massDistanceGF+} and \ref{fig:massDistanceGF-}.
The distance to the source was assessed using archival photometric data, because the parallax $\varpi$ in the published Gaia catalogues for this source is negative.
The estimated distance to the source varies between 14.5~kpc and 23~kpc, with a median value of 18.2~kpc.
The estimated mass of the lens is $\ml = 2.65^{+5.09}_{-1.48} M_\odot,$ and the distance is $\dl = 2.84^{+1.94}_{-1.64}$~kpc for the solution with $\unod>0$ (\mbox{GF+}).
For the solution with $u_0<0$ (\mbox{GF-}), the lens weighs $\ml = 1.71^{+3.78}_{-1.06} M_\odot$, while the distance to the lens is $\dl = 2.66^{+1.97}_{-1.56}$~kpc.
Because the event is dim and the blending parameter suggests that the majority of the light comes from the source, we propose that the lens is a dark remnant candidate, possibly a massive neutron star, or a mass-gap black hole.
We conclude that it is impossible to confirm the hypothesis of the lens being a dark remnant right now, but follow-up observations in the future could resolve this issue.
If the event was caused by a massive lens, we will be able see the signature of microlensing in the astrometric time series published with the final Gaia Data Release~(2024).


\begin{acknowledgements}
The authors would like to thank Dr Radek Poleski for discussion and his comments during the creation of this work.
We would like to thank the referee for their comments, which helped to improve this paper.
We acknowledge ESA Gaia, DPAC and the Photometric Science Alerts Team (http://gsaweb.ast.cam.ac.uk/alerts).
This work was supported from the Polish NCN grants: Harmonia No. 2018/30/M/ST9/00311, Daina No. 2017/27/L/ST9/03221, 
MNiSW grant DIR/WK/2018/12 and
NCBiR grant within POWER program nr POWR.03.02.00-00-l001/16-00.
We acknowledge Research Council of Lithuania grant No S-LL-19-2 and
European Commission's H2020 OPTICON grant No. 730890 as well as OPTICON RadioNet Pilot grant No. 101004719.
YT acknowledges the support of DFG priority program SPP 1992 “Exploring the Diversity of Extrasolar Planets” (TS 356/3-1).
This work is partly supported by JSPS KAKENHI Grant Number JP18H05439, and the Astrobiology Center of National Institutes of Natural Sciences (NINS) (Grant Number AB031010).
\end{acknowledgements}

%
%
\bibliographystyle{aa}
\bibliography{Gaia18cbf}

\begin{thebibliography}{84}
\expandafter\ifx\csname natexlab\endcsname\relax\def\natexlab#1{#1}\fi

\bibitem[{{Abbott} {et~al.}(2020){Abbott}, {Abbott}, {Abraham}, {Acernese},
  {Ackley}, {Adams}, {Adhikari}, {Adya}, {Affeldt}, {Agathos}, {Agatsuma},
  {Aggarwal}, {Aguiar}, {Aich}, {Aiello}, {Ain}, {Ajith}, {Akcay}, {Allen},
  {Allocca}, {Altin}, {Amato}, {Anand}, {Ananyeva}, {Anderson}, {Anderson},
  {Angelova}, {Ansoldi}, {Antier}, {Appert}, {Arai}, {Araya}, {Areeda},
  {Ar{\`e}ne}, {Arnaud}, {Aronson}, {Arun}, {Asali}, {Ascenzi}, {Ashton},
  {Aston}, {Astone}, {Aubin}, {Aufmuth}, {AultONeal}, {Austin}, {Avendano},
  {Babak}, {Bacon}, {Badaracco}, {Bader}, {Bae}, {Baer}, {Baird}, {Baldaccini},
  {Ballardin}, {Ballmer}, {Bals}, {Balsamo}, {Baltus}, {Banagiri}, {Bankar},
  {Bankar}, {Barayoga}, {Barbieri}, {Barish}, {Barker}, {Barkett}, {Barneo},
  {Barone}, {Barr}, {Barsotti}, {Barsuglia}, {Barta}, {Bartlett}, {Bartos},
  {Bassiri}, {Basti}, {Bawaj}, {Bayley}, {Bazzan}, {B{\'e}csy}, {Bejger},
  {Belahcene}, {Bell}, {Beniwal}, {Benjamin}, {Benkel}, {Bentley}, {Bergamin},
  {Berger}, {Bergmann}, {Bernuzzi}, {Berry}, {Bersanetti}, {Bertolini},
  {Betzwieser}, {Bhandare}, {Bhandari}, {Bidler}, {Biggs}, {Bilenko},
  {Billingsley}, {Birney}, {Birnholtz}, {Biscans}, {Bischi}, {Biscoveanu},
  {Bisht}, {Bissenbayeva}, {Bitossi}, {Bizouard}, {Blackburn}, {Blackman},
  {Blair}, {Blair}, {Blair}, {Bobba}, {Bode}, {Boer}, {Boetzel}, {Bogaert},
  {Bondu}, {Bonilla}, {Bonnand}, {Booker}, {Boom}, {Bork}, {Boschi}, {Bose},
  {Bossilkov}, {Bosveld}, {Bouffanais}, {Bozzi}, {Bradaschia}, {Brady},
  {Bramley}, {Branchesi}, {Brau}, {Breschi}, {Briant}, {Briggs}, {Brighenti},
  {Brillet}, {Brinkmann}, {Brito}, {Brockill}, {Brooks}, {Brooks}, {Brown},
  {Brunett}, {Bruno}, {Bruntz}, {Buikema}, {Bulik}, {Bulten}, {Buonanno},
  {Buskulic}, {Byer}, {Cabero}, {Cadonati}, {Cagnoli}, {Cahillane}, {Bustillo},
  {Callaghan}, {Callister}, {Calloni}, {Camp}, {Canepa}, {Cannon}, {Cao},
  {Cao}, {Carapella}, {Carbognani}, {Caride}, {Carney}, {Carullo}, {Diaz},
  {Casentini}, {Casta{\~n}eda}, {Caudill}, {Cavagli{\`a}}, {Cavalier},
  {Cavalieri}, {Cella}, {Cerd{\'a}-Dur{\'a}n}, {Cesarini}, {Chaibi},
  {Chakravarti}, {Chan}, {Chan}, {Chao}, {Charlton}, {Chase},
  {Chassande-Mottin}, {Chatterjee}, {Chaturvedi}, {Chatziioannou}, {Chen},
  {Chen}, {Chen}, {Cheng}, {Cheong}, {Chia}, {Chiadini}, {Chierici},
  {Chincarini}, {Chiummo}, {Cho}, {Cho}, {Cho}, {Christensen}, {Chu}, {Chua},
  {Chung}, {Chung}, {Ciani}, {Ciecielag}, {Cie{\'s}lar}, {Ciobanu}, {Ciolfi},
  {Cipriano}, {Cirone}, {Clara}, {Clark}, {Clearwater}, {Clesse}, {Cleva},
  {Coccia}, {Cohadon}, {Cohen}, {Colleoni}, {Collette}, {Collins}, {Colpi},
  {Constancio}, {Conti}, {Cooper}, {Corban}, {Corbitt}, {Cordero-Carri{\'o}n},
  {Corezzi}, {Corley}, {Cornish}, {Corre}, {Corsi}, {Cortese}, {Costa},
  {Cotesta}, {Coughlin}, {Coughlin}, {Coulon}, {Countryman}, {Couvares},
  {Covas}, {Coward}, {Cowart}, {Coyne}, {Coyne}, {Creighton}, {Creighton},
  {Cripe}, {Croquette}, {Crowder}, {Cudell}, {Cullen}, {Cumming}, {Cummings},
  {Cunningham}, {Cuoco}, {Curylo}, {Canton}, {D{\'a}lya}, {Dana},
  {Daneshgaran-Bajastani}, {D'Angelo}, {Danilishin}, {D'Antonio}, {Danzmann},
  {Darsow-Fromm}, {Dasgupta}, {Datrier}, {Dattilo}, {Dave}, {Davier}, {Davies},
  {Davis}, {Daw}, {DeBra}, {Deenadayalan}, {Degallaix}, {De Laurentis},
  {Del{\'e}glise}, {Delfavero}, {De Lillo}, {Del Pozzo}, {DeMarchi},
  {D'Emilio}, {Demos}, {Dent}, {De Pietri}, {De Rosa}, {De Rossi}, {DeSalvo},
  {de Varona}, {Dhurandhar}, {D{\'\i}az}, {Diaz-Ortiz}, {Dietrich}, {Di Fiore},
  {Di Fronzo}, {Di Giorgio}, {Di Giovanni}, {Di Giovanni}, {Di Girolamo}, {Di
  Lieto}, {Ding}, {Di Pace}, {Di Palma}, {Di Renzo}, {Divakarla}, {Dmitriev},
  {Doctor}, {Donovan}, {Dooley}, {Doravari}, {Dorrington}, {Downes}, {Drago},
  {Driggers}, {Du}, {Ducoin}, {Dupej}, {Durante}, {D'Urso}, {Dwyer}, {Easter},
  {Eddolls}, {Edelman}, {Edo}, {Edy}, {Effler}, {Ehrens}, {Eichholz},
  {Eikenberry}, {Eisenmann}, {Eisenstein}, {Ejlli}, {Errico}, {Essick},
  {Estelles}, {Estevez}, {Etienne}, {Etzel}, {Evans}, {Evans}, {Ewing},
  {Fafone}, {Fairhurst}, {Fan}, {Farinon}, {Farr}, {Farr}, {Fauchon-Jones},
  {Favata}, {Fays}, {Fazio}, {Feicht}, {Fejer}, {Feng}, {Fenyvesi}, {Ferguson},
  {Fernandez-Galiana}, {Ferrante}, {Ferreira}, {Ferreira}, {Fidecaro}, {Fiori},
  {Fiorucci}, {Fishbach}, {Fisher}, {Fittipaldi}, {Fitz-Axen}, {Fiumara},
  {Flaminio}, {Floden}, {Flynn}, {Fong}, {Font}, {Forsyth}, {Fournier},
  {Frasca}, {Frasconi}, {Frei}, {Freise}, {Frey}, {Frey}, {Fritschel},
  {Frolov}, {Fronz{\`e}}, {Fulda}, {Fyffe}, {Gabbard}, {Gadre}, {Gaebel},
  {Gair}, {Galaudage}, {Ganapathy}, {Ganguly}, {Gaonkar},
  {Garc{\'\i}a-Quir{\'o}s}, {Garufi}, {Gateley}, {Gaudio}, {Gayathri}, {Gemme},
  {Genin}, {Gennai}, {George}, {George}, {Gergely}, {Ghonge}, {Ghosh}, {Ghosh},
  {Ghosh}, {Giacomazzo}, {Giaime}, {Giardina}, {Gibson}, {Gier}, {Gill},
  {Glanzer}, {Gniesmer}, {Godwin}, {Goetz}, {Goetz}, {Gohlke}, {Goncharov},
  {Gonz{\'a}lez}, {Gopakumar}, {Gossan}, {Gosselin}, {Gouaty}, {Grace},
  {Grado}, {Granata}, {Grant}, {Gras}, {Grassia}, {Gray}, {Gray}, {Greco},
  {Green}, {Green}, {Gretarsson}, {Griggs}, {Grignani}, {Grimaldi}, {Grimm},
  {Grote}, {Grunewald}, {Gruning}, {Guidi}, {Guimaraes}, {Guix{\'e}}, {Gulati},
  {Guo}, {Gupta}, {Gupta}, {Gupta}, {Gustafson}, {Gustafson}, {Haegel},
  {Halim}, {Hall}, {Hamilton}, {Hammond}, {Haney}, {Hanke}, {Hanks}, {Hanna},
  {Hannam}, {Hannuksela}, {Hansen}, {Hanson}, {Harder}, {Hardwick}, {Haris},
  {Harms}, {Harry}, {Harry}, {Hasskew}, {Haster}, {Haughian}, {Hayes}, {Healy},
  {Heidmann}, {Heintze}, {Heinze}, {Heitmann}, {Hellman}, {Hello}, {Hemming},
  {Hendry}, {Heng}, {Hennes}, {Hennig}, {Heurs}, {Hild}, {Hinderer}, {Hoback},
  {Hochheim}, {Hofgard}, {Hofman}, {Holgado}, {Holland}, {Holt}, {Holz},
  {Hopkins}, {Horst}, {Hough}, {Howell}, {Hoy}, {Huang}, {H{\"u}bner},
  {Huerta}, {Huet}, {Hughey}, {Hui}, {Husa}, {Huttner}, {Huxford},
  {Huynh-Dinh}, {Idzkowski}, {Iess}, {Inchauspe}, {Ingram}, {Intini}, {Isac},
  {Isi}, {Iyer}, {Jacqmin}, {Jadhav}, {Jadhav}, {James}, {Jani}, {Janthalur},
  {Jaranowski}, {Jariwala}, {Jaume}, {Jenkins}, {Jiang}, {Johns},
  {Johnson-McDaniel}, {Jones}, {Jones}, {Jones}, {Jones}, {Jones}, {Jonker},
  {Ju}, {Junker}, {Kalaghatgi}, {Kalogera}, {Kamai}, {Kandhasamy}, {Kang},
  {Kanner}, {Kapadia}, {Karki}, {Kashyap}, {Kasprzack}, {Kastaun},
  {Katsanevas}, {Katsavounidis}, {Katzman}, {Kaufer}, {Kawabe},
  {K{\'e}f{\'e}lian}, {Keitel}, {Keivani}, {Kennedy}, {Key}, {Khadka},
  {Khalili}, {Khan}, {Khan}, {Khan}, {Khazanov}, {Khetan}, {Khursheed},
  {Kijbunchoo}, {Kim}, {Kim}, {Kim}, {Kim}, {Kim}, {Kim}, {Kim}, {Kimball},
  {King}, {Kinley-Hanlon}, {Kirchhoff}, {Kissel}, {Kleybolte}, {Klimenko},
  {Knowles}, {Knyazev}, {Koch}, {Koehlenbeck}, {Koekoek}, {Koley},
  {Kondrashov}, {Kontos}, {Koper}, {Korobko}, {Korth}, {Kovalam}, {Kozak},
  {Kringel}, {Krishnendu}, {Kr{\'o}lak}, {Krupinski}, {Kuehn}, {Kumar},
  {Kumar}, {Kumar}, {Kumar}, {Kumar}, {Kuo}, {Kutynia}, {Lackey}, {Laghi},
  {Lalande}, {Lam}, {Lamberts}, {Landry}, {Landry}, {Lane}, {Lang}, {Lange},
  {Lantz}, {Lanza}, {La Rosa}, {Lartaux-Vollard}, {Lasky}, {Laxen},
  {Lazzarini}, {Lazzaro}, {Leaci}, {Leavey}, {Lecoeuche}, {Lee}, {Lee}, {Lee},
  {Lee}, {Lee}, {Lehmann}, {Leroy}, {Letendre}, {Levin}, {Li}, {Li}, {li},
  {Li}, {Li}, {Linde}, {Linker}, {Linley}, {Littenberg}, {Liu}, {Liu},
  {Llorens-Monteagudo}, {Lo}, {Lockwood}, {London}, {Longo}, {Lorenzini},
  {Loriette}, {Lormand}, {Losurdo}, {Lough}, {Lousto}, {Lovelace}, {L{\"u}ck},
  {Lumaca}, {Lundgren}, {Ma}, {Macas}, {Macfoy}, {MacInnis}, {Macleod},
  {MacMillan}, {Macquet}, {Hernandez}, {Maga{\~n}a-Sandoval}, {Magee},
  {Majorana}, {Maksimovic}, {Malik}, {Man}, {Mandic}, {Mangano}, {Mansell},
  {Manske}, {Mantovani}, {Mapelli}, {Marchesoni}, {Marion}, {M{\'a}rka},
  {M{\'a}rka}, {Markakis}, {Markosyan}, {Markowitz}, {Maros}, {Marquina},
  {Marsat}, {Martelli}, {Martin}, {Martin}, {Martinez}, {Martynov},
  {Masalehdan}, {Mason}, {Massera}, {Masserot}, {Massinger}, {Masso-Reid},
  {Mastrogiovanni}, {Matas}, {Matichard}, {Mavalvala}, {Maynard}, {McCann},
  {McCarthy}, {McClelland}, {McCormick}, {McCuller}, {McGuire}, {McIsaac},
  {McIver}, {McManus}, {McRae}, {McWilliams}, {Meacher}, {Meadors}, {Mehmet},
  {Mehta}, {Villa}, {Melatos}, {Mendell}, {Mercer}, {Mereni}, {Merfeld},
  {Merilh}, {Merritt}, {Merzougui}, {Meshkov}, {Messenger}, {Messick},
  {Metzdorff}, {Meyers}, {Meylahn}, {Mhaske}, {Miani}, {Miao}, {Michaloliakos},
  {Michel}, {Middleton}, {Milano}, {Miller}, {Millhouse}, {Mills}, {Milotti},
  {Milovich-Goff}, {Minazzoli}, {Minenkov}, {Mishkin}, {Mishra}, {Mistry},
  {Mitra}, {Mitrofanov}, {Mitselmakher}, {Mittleman}, {Mo}, {Mogushi},
  {Mohapatra}, {Mohite}, {Molina-Ruiz}, {Mondin}, {Montani}, {Moore}, {Moraru},
  {Morawski}, {Moreno}, {Morisaki}, {Mours}, {Mow-Lowry}, {Mozzon},
  {Muciaccia}, {Mukherjee}, {Mukherjee}, {Mukherjee}, {Mukherjee}, {Mukund},
  {Mullavey}, {Munch}, {Mu{\~n}iz}, {Murray}, {Nagar}, {Nardecchia},
  {Naticchioni}, {Nayak}, {Neil}, {Neilson}, {Nelemans}, {Nelson}, {Nery},
  {Neunzert}, {Ng}, {Ng}, {Nguyen}, {Nguyen}, {Nichols}, {Nichols}, {Nissanke},
  {Nocera}, {Noh}, {North}, {Nothard}, {Nuttall}, {Oberling}, {O'Brien},
  {Oganesyan}, {Ogin}, {Oh}, {Oh}, {Ohme}, {Ohta}, {Okada}, {Oliver},
  {Olivetto}, {Oppermann}, {Oram}, {O'Reilly}, {Ormiston}, {Ortega},
  {O'Shaughnessy}, {Ossokine}, {Osthelder}, {Ottaway}, {Overmier}, {Owen},
  {Pace}, {Pagano}, {Page}, {Pagliaroli}, {Pai}, {Pai}, {Palamos}, {Palashov},
  {Palomba}, {Pan}, {Panda}, {Pang}, {Pankow}, {Pannarale}, {Pant}, {Paoletti},
  {Paoli}, {Parida}, {Parker}, {Pascucci}, {Pasqualetti}, {Passaquieti},
  {Passuello}, {Patricelli}, {Payne}, {Pearlstone}, {Pechsiri}, {Pedersen},
  {Pedraza}, {Pele}, {Penn}, {Perego}, {Perez}, {P{\'e}rigois}, {Perreca},
  {Perri{\`e}s}, {Petermann}, {Pfeiffer}, {Phelps}, {Phukon}, {Piccinni},
  {Pichot}, {Piendibene}, {Piergiovanni}, {Pierro}, {Pillant}, {Pinard},
  {Pinto}, {Piotrzkowski}, {Pirello}, {Pitkin}, {Plastino}, {Poggiani}, {Pong},
  {Ponrathnam}, {Popolizio}, {Porter}, {Powell}, {Prajapati}, {Prasai},
  {Prasanna}, {Pratten}, {Prestegard}, {Principe}, {Prodi}, {Prokhorov},
  {Punturo}, {Puppo}, {P{\"u}rrer}, {Qi}, {Quetschke}, {Quinonez}, {Raab},
  {Raaijmakers}, {Radkins}, {Radulesco}, {Raffai}, {Rafferty}, {Raja}, {Rajan},
  {Rajbhandari}, {Rakhmanov}, {Ramirez}, {Ramos-Buades}, {Rana}, {Rao},
  {Rapagnani}, {Raymond}, {Razzano}, {Read}, {Regimbau}, {Rei}, {Reid},
  {Reitze}, {Rettegno}, {Ricci}, {Richardson}, {Richardson}, {Ricker},
  {Riemenschneider}, {Riles}, {Rizzo}, {Robertson}, {Robinet}, {Rocchi},
  {Rodriguez-Soto}, {Rolland}, {Rollins}, {Roma}, {Romanelli}, {Romano},
  {Romel}, {Romero-Shaw}, {Romie}, {Rose}, {Rose}, {Rose}, {Rosi{\'n}ska},
  {Rosofsky}, {Ross}, {Rowan}, {Rowlinson}, {Roy}, {Roy}, {Roy}, {Ruggi},
  {Rutins}, {Ryan}, {Sachdev}, {Sadecki}, {Sakellariadou}, {Salafia},
  {Salconi}, {Saleem}, {Salemi}, {Samajdar}, {Sanchez}, {Sanchez},
  {Sanchis-Gual}, {Sanders}, {Santiago}, {Santos}, {Sarin}, {Sassolas},
  {Sathyaprakash}, {Sauter}, {Savage}, {Savant}, {Sawant}, {Sayah}, {Schaetzl},
  {Schale}, {Scheel}, {Scheuer}, {Schmidt}, {Schnabel}, {Schofield},
  {Sch{\"o}nbeck}, {Schreiber}, {Schulte}, {Schutz}, {Schwarm}, {Schwartz},
  {Scott}, {Scott}, {Seidel}, {Sellers}, {Sengupta}, {Sennett}, {Sentenac},
  {Sequino}, {Sergeev}, {Setyawati}, {Shaddock}, {Shaffer}, {Shahriar},
  {Sharma}, {Sharma}, {Shawhan}, {Shen}, {Shikauchi}, {Shink}, {Shoemaker},
  {Shoemaker}, {Shukla}, {ShyamSundar}, {Siellez}, {Sieniawska}, {Sigg},
  {Singer}, {Singh}, {Singh}, {Singha}, {Singhal}, {Sintes}, {Sipala},
  {Skliris}, {Slagmolen}, {Slaven-Blair}, {Smetana}, {Smith}, {Smith},
  {Somala}, {Son}, {Soni}, {Sorazu}, {Sordini}, {Sorrentino}, {Souradeep},
  {Sowell}, {Spencer}, {Spera}, {Srivastava}, {Srivastava}, {Staats},
  {Stachie}, {Standke}, {Steer}, {Steinhoff}, {Steinke}, {Steinlechner},
  {Steinlechner}, {Steinmeyer}, {Stevenson}, {Stocks}, {Stops}, {Stover},
  {Strain}, {Stratta}, {Strunk}, {Sturani}, {Stuver}, {Sudhagar}, {Sudhir},
  {Summerscales}, {Sun}, {Sunil}, {Sur}, {Suresh}, {Sutton}, {Swinkels},
  {Szczepa{\'n}czyk}, {Tacca}, {Tait}, {Talbot}, {Tanasijczuk}, {Tanner},
  {Tao}, {T{\'a}pai}, {Tapia}, {San Martin}, {Tasson}, {Taylor}, {Tenorio},
  {Terkowski}, {Thirugnanasambandam}, {Thomas}, {Thomas}, {Thompson},
  {Thondapu}, {Thorne}, {Thrane}, {Tinsman}, {Saravanan}, {Tiwari}, {Tiwari},
  {Tiwari}, {Toland}, {Tonelli}, {Tornasi}, {Torres-Forn{\'e}}, {Torrie},
  {Tosta e Melo}, {T{\"o}yr{\"a}}, {Trail}, {Travasso}, {Traylor}, {Tringali},
  {Tripathee}, {Trovato}, {Trudeau}, {Tsang}, {Tse}, {Tso}, {Tsukada}, {Tsuna},
  {Tsutsui}, {Turconi}, {Ubhi}, {Ueno}, {Ugolini}, {Unnikrishnan}, {Urban},
  {Usman}, {Utina}, {Vahlbruch}, {Vajente}, {Valdes}, {Valentini}, {van Bakel},
  {van Beuzekom}, {van den Brand}, {Van Den Broeck}, {Vander-Hyde}, {van der
  Schaaf}, {Van Heijningen}, {van Veggel}, {Vardaro}, {Varma}, {Vass},
  {Vas{\'u}th}, {Vecchio}, {Vedovato}, {Veitch}, {Veitch}, {Venkateswara},
  {Venugopalan}, {Verkindt}, {Veske}, {Vetrano}, {Vicer{\'e}}, {Viets},
  {Vinciguerra}, {Vine}, {Vinet}, {Vitale}, {Vivanco}, {Vo}, {Vocca},
  {Vorvick}, {Vyatchanin}, {Wade}, {Wade}, {Wade}, {Walet}, {Walker},
  {Wallace}, {Wallace}, {Walsh}, {Wang}, {Wang}, {Wang}, {Ward}, {Warden},
  {Warner}, {Was}, {Watchi}, {Weaver}, {Wei}, {Weinert}, {Weinstein}, {Weiss},
  {Wellmann}, {Wen}, {We{\ss}els}, {Westhouse}, {Wette}, {Whelan}, {Whiting},
  {Whittle}, {Wilken}, {Williams}, {Willis}, {Willke}, {Winkler}, {Wipf},
  {Wittel}, {Woan}, {Woehler}, {Wofford}, {Wong}, {Wright}, {Wu}, {Wysocki},
  {Xiao}, {Yamamoto}, {Yang}, {Yang}, {Yang}, {Yap}, {Yazback}, {Yeeles}, {Yu},
  {Yu}, {Yuen}, {Zadro{\.z}ny}, {Zadro{\.z}ny}, {Zanolin}, {Zelenova},
  {Zendri}, {Zevin}, {Zhang}, {Zhang}, {Zhang}, {Zhao}, {Zhao}, {Zhou}, {Zhou},
  {Zhu}, {Zimmerman}, {Zucker}, {Zweizig}, {LIGO Scientific Collaboration}, \&
  {Virgo Collaboration}}]{2020AbbotGWmassgap}
{Abbott}, R., {Abbott}, T.~D., {Abraham}, S., {et~al.} 2020, \apjl, 896, L44

\bibitem[{{Abe} {et~al.}(1997){Abe}, {Allen}, {Banks}, {Bond}, {Carter},
  {Dodd}, {Fujimoto}, {Yayashida}, {Hearnshaw}, {Honda}, {Jugaku}, {Kabe},
  {Kobayashi}, {Kilmartin}, {Kitamura}, {Love}, {Matsubara}, {Mityamoto},
  {Muraki}, {Nakamura}, {Pennycook}, {Pipe}, {Reid}, {Sato}, {Sato}, {Saito},
  {Sekiguchi}, {Sullivan}, {Watase}, {Yock}, {Yanagisawa}, \&
  {Yoshizawa}}]{1997vsar.conf...75A}
{Abe}, F., {Allen}, W., {Banks}, T., {et~al.} 1997, in Variables Stars and the
  Astrophysical Returns of the Microlensing Surveys, ed. R.~{Ferlet}, J.-P.
  {Maillard}, \& B.~{Raban}, 75

\bibitem[{{Abrams} \& {Takada}(2020)}]{2020Abrams}
{Abrams}, N.~S. \& {Takada}, M. 2020, \apj, 905, 121

\bibitem[{{Alcock} {et~al.}(1995){Alcock}, {Allsman}, {Alves}, {Axelrod},
  {Bennett}, {Cook}, {Freeman}, {Griest}, {Guern}, {Lehner}, {Marshall},
  {Peterson}, {Pratt}, {Quinn}, {Rodgers}, {Stubbs}, \&
  {Sutherland}}]{1995ApJ...454L.125A}
{Alcock}, C., {Allsman}, R.~A., {Alves}, D., {et~al.} 1995, \apjl, 454, L125

\bibitem[{{Alcock} {et~al.}(1992){Alcock}, {Axelrod}, {Bennett}, {Cook},
  {Park}, {Griest}, {Perlmutter}, {Stubbs}, {Freeman}, {Peterson}, {Quinn}, \&
  {Rodgers}}]{1992ASPC...34..193A}
{Alcock}, C., {Axelrod}, T.~S., {Bennett}, D.~P., {et~al.} 1992, in
  Astronomical Society of the Pacific Conference Series, Vol. 103, Robotic
  Telescopes in the 1990s, ed. A.~V. {Filippenko}, 193--202

\bibitem[{{Amaro-Seoane} {et~al.}(2017){Amaro-Seoane}, {Audley}, {Babak},
  {Baker}, {Barausse}, {Bender}, {Berti}, {Binetruy}, {Born}, {Bortoluzzi},
  {Camp}, {Caprini}, {Cardoso}, {Colpi}, {Conklin}, {Cornish}, {Cutler},
  {Danzmann}, {Dolesi}, {Ferraioli}, {Ferroni}, {Fitzsimons}, {Gair}, {Gesa
  Bote}, {Giardini}, {Gibert}, {Grimani}, {Halloin}, {Heinzel}, {Hertog},
  {Hewitson}, {Holley-Bockelmann}, {Hollington}, {Hueller}, {Inchauspe},
  {Jetzer}, {Karnesis}, {Killow}, {Klein}, {Klipstein}, {Korsakova}, {Larson},
  {Livas}, {Lloro}, {Man}, {Mance}, {Martino}, {Mateos}, {McKenzie},
  {McWilliams}, {Miller}, {Mueller}, {Nardini}, {Nelemans}, {Nofrarias},
  {Petiteau}, {Pivato}, {Plagnol}, {Porter}, {Reiche}, {Robertson},
  {Robertson}, {Rossi}, {Russano}, {Schutz}, {Sesana}, {Shoemaker}, {Slutsky},
  {Sopuerta}, {Sumner}, {Tamanini}, {Thorpe}, {Troebs}, {Vallisneri},
  {Vecchio}, {Vetrugno}, {Vitale}, {Volonteri}, {Wanner}, {Ward}, {Wass},
  {Weber}, {Ziemer}, \& {Zweifel}}]{2017LISA}
{Amaro-Seoane}, P., {Audley}, H., {Babak}, S., {et~al.} 2017, arXiv e-prints,
  arXiv:1702.00786

\bibitem[{{An} {et~al.}(2002){An}, {Albrow}, {Beaulieu}, {Caldwell}, {DePoy},
  {Dominik}, {Gaudi}, {Gould}, {Greenhill}, {Hill}, {Kane}, {Martin},
  {Menzies}, {Pogge}, {Pollard}, {Sackett}, {Sahu}, {Vermaak}, {Watson}, \&
  {Williams}}]{2002An}
{An}, J.~H., {Albrow}, M.~D., {Beaulieu}, J.~P., {et~al.} 2002, \apj, 572, 521

\bibitem[{{Aubourg} {et~al.}(1993){Aubourg}, {Bareyre}, {Brehin}, {Gros},
  {Lachieze-Rey}, {Laurent}, {Lesquoy}, {Magneville}, {Milsztajn}, {Moscoso},
  {Queinnec}, {Rich}, {Spiro}, {Vigroux}, {Zylberajch}, {Ansari}, {Cavalier},
  {Moniez}, {Beaulieu}, {Ferlet}, {Grison}, {Vidal-Madjar}, {Guibert},
  {Moreau}, {Tajahmady}, {Maurice}, {Prevot}, \& {Gry}}]{1993Msngr..72...20A}
{Aubourg}, E., {Bareyre}, P., {Brehin}, S., {et~al.} 1993, The Messenger, 72,
  20

\bibitem[{{Bailer-Jones} {et~al.}(2021){Bailer-Jones}, {Rybizki}, {Fouesneau},
  {Demleitner}, \& {Andrae}}]{2021Bailer-JonesEDR3}
{Bailer-Jones}, C.~A.~L., {Rybizki}, J., {Fouesneau}, M., {Demleitner}, M., \&
  {Andrae}, R. 2021, \aj, 161, 147

\bibitem[{{Bailer-Jones} {et~al.}(2018){Bailer-Jones}, {Rybizki}, {Fouesneau},
  {Mantelet}, \& {Andrae}}]{BailerJones}
{Bailer-Jones}, C.~A.~L., {Rybizki}, J., {Fouesneau}, M., {Mantelet}, G., \&
  {Andrae}, R. 2018, \aj, 156, 58

\bibitem[{{Bastian} {et~al.}(2010){Bastian}, {Covey}, \& {Meyer}}]{BastianIMF}
{Bastian}, N., {Covey}, K.~R., \& {Meyer}, M.~R. 2010, \araa, 48, 339

\bibitem[{{Belokurov} \& {Evans}(2002)}]{2002MNRAS.331..649B}
{Belokurov}, V.~A. \& {Evans}, N.~W. 2002, \mnras, 331, 649

\bibitem[{{Bennett} {et~al.}(1993){Bennett}, {Akerlof}, {Alcock}, {Allsman},
  {Axelrod}, {Cook}, {Freeman}, {Griest}, {Marshall}, {Park}, {Perlmutter},
  {Peterson}, {Quinn}, {Rodgers}, {Stubbs}, \&
  {Sutherland}}]{1993NYASA.688..612B}
{Bennett}, D.~P., {Akerlof}, C., {Alcock}, C., {et~al.} 1993, in Texas/PASCOS
  '92: Relativistic Astrophysics and Particle Cosmology, ed. C.~W. {Akerlof} \&
  M.~A. {Srednicki}, Vol. 688, 612

\bibitem[{{Bennett} {et~al.}(2002){Bennett}, {Becker}, {Quinn}, {Tomaney},
  {Alcock}, {Allsman}, {Alves}, {Axelrod}, {Calitz}, {Cook}, {Drake},
  {Fragile}, {Freeman}, {Geha}, {Griest}, {Johnson}, {Keller}, {Laws},
  {Lehner}, {Marshall}, {Minniti}, {Nelson}, {Peterson}, {Popowski}, {Pratt},
  {Quinn}, {Rhie}, {Stubbs}, {Sutherland}, {Vandehei}, {Welch}, {MACHO
  Collaboration}, \& {MPS Collaboration}}]{2002Bennett}
{Bennett}, D.~P., {Becker}, A.~C., {Quinn}, J.~L., {et~al.} 2002, \apj, 579,
  639

\bibitem[{{Blagorodnova} {et~al.}(2016){Blagorodnova}, {Van Velzen},
  {Harrison}, {Koposov}, {Mattila}, {Campbell}, {Walton}, \&
  {Wyrzykowski}}]{2016MNRAS.455..603B}
{Blagorodnova}, N., {Van Velzen}, S., {Harrison}, D.~L., {et~al.} 2016, \mnras,
  455, 603

\bibitem[{{Brown} {et~al.}(2013){Brown}, {Baliber}, {Bianco}, {Bowman},
  {Burleson}, {Conway}, {Crellin}, {Depagne}, {De Vera}, {Dilday}, {Dragomir},
  {Dubberley}, {Eastman}, {Elphick}, {Falarski}, {Foale}, {Ford}, {Fulton},
  {Garza}, {Gomez}, {Graham}, {Greene}, {Haldeman}, {Hawkins}, {Haworth},
  {Haynes}, {Hidas}, {Hjelstrom}, {Howell}, {Hygelund}, {Lister}, {Lobdill},
  {Martinez}, {Mullins}, {Norbury}, {Parrent}, {Paulson}, {Petry}, {Pickles},
  {Posner}, {Rosing}, {Ross}, {Sand}, {Saunders}, {Shobbrook}, {Shporer},
  {Street}, {Thomas}, {Tsapras}, {Tufts}, {Valenti}, {Vander Horst}, {Walker},
  {White}, \& {Willis}}]{LCONetw}
{Brown}, T.~M., {Baliber}, N., {Bianco}, F.~B., {et~al.} 2013, \pasp, 125, 1031

\bibitem[{{Capitanio} {et~al.}(2017){Capitanio}, {Lallement}, {Vergely},
  {Elyajouri}, \& {Monreal-Ibero}}]{Reddening2}
{Capitanio}, L., {Lallement}, R., {Vergely}, J.~L., {Elyajouri}, M., \&
  {Monreal-Ibero}, A. 2017, \aap, 606, A65

\bibitem[{{Cassan} {et~al.}(2021){Cassan}, {Ranc}, {Absil}, {Wyrzykowski},
  {Rybicki}, {Bachelet}, {Le Bouquin}, {Hundertmark}, {Street}, {Surdej},
  {Tsapras}, {Wambsganss}, \& {Wertz}}]{CassanGaia19bld}
{Cassan}, A., {Ranc}, C., {Absil}, O., {et~al.} 2021, Nature Astronomy

\bibitem[{{Dong} {et~al.}(2019){Dong}, {M{\'e}rand}, {Delplancke-Str{\"o}bele},
  {Gould}, {Chen}, {Post}, {Kochanek}, {Stanek}, {Christie}, {Mutel},
  {Natusch}, {Holoien}, {Prieto}, {Shappee}, \& {Thompson}}]{2019DongVLTI}
{Dong}, S., {M{\'e}rand}, A., {Delplancke-Str{\"o}bele}, F., {et~al.} 2019,
  \apj, 871, 70

\bibitem[{{Dong} {et~al.}(2007){Dong}, {Udalski}, {Gould}, {Reach}, {Christie},
  {Boden}, {Bennett}, {Fazio}, {Griest}, {Szyma{\'n}ski}, {Kubiak},
  {Soszy{\'n}ski}, {Pietrzy{\'n}ski}, {Szewczyk}, {Wyrzykowski}, {Ulaczyk},
  {Wieckowski}, {Paczy{\'n}ski}, {DePoy}, {Pogge}, {Preston}, {Thompson}, \&
  {Patten}}]{2007ApJ...664..862D}
{Dong}, S., {Udalski}, A., {Gould}, A., {et~al.} 2007, \apj, 664, 862

\bibitem[{{Einstein}(1936)}]{1936Einstein}
{Einstein}, A. 1936, Science, 84, 506

\bibitem[{{Evans} {et~al.}(2018){Evans}, {Riello}, {De Angeli}, {Carrasco},
  {Montegriffo}, {Fabricius}, {Jordi}, {Palaversa}, {Diener}, {Busso},
  {Cacciari}, {van Leeuwen}, {Burgess}, {Davidson}, {Harrison}, {Hodgkin},
  {Pancino}, {Richards}, {Altavilla}, {Balaguer-N{\'u}{\~n}ez}, {Barstow},
  {Bellazzini}, {Brown}, {Castellani}, {Cocozza}, {De Luise}, {Delgado},
  {Ducourant}, {Galleti}, {Gilmore}, {Giuffrida}, {Holl}, {Kewley}, {Koposov},
  {Marinoni}, {Marrese}, {Osborne}, {Piersimoni}, {Portell}, {Pulone},
  {Ragaini}, {Sanna}, {Terrett}, {Walton}, {Wevers}, \&
  {Wyrzykowski}}]{2018EvansPhoto}
{Evans}, D.~W., {Riello}, M., {De Angeli}, F., {et~al.} 2018, \aap, 616, A4

\bibitem[{{Farr} {et~al.}(2011){Farr}, {Sravan}, {Cantrell}, {Kreidberg},
  {Bailyn}, {Mandel}, \& {Kalogera}}]{2011FarrBHmass}
{Farr}, W.~M., {Sravan}, N., {Cantrell}, A., {et~al.} 2011, \apj, 741, 103

\bibitem[{Foreman-Mackey(2016)}]{corner}
Foreman-Mackey, D. 2016, The Journal of Open Source Software, 1, 24

\bibitem[{{Foreman-Mackey} {et~al.}(2013){Foreman-Mackey}, {Hogg}, {Lang}, \&
  {Goodman}}]{2013Emcee}
{Foreman-Mackey}, D., {Hogg}, D.~W., {Lang}, D., \& {Goodman}, J. 2013, \pasp,
  125, 306

\bibitem[{{Gaia Collaboration} {et~al.}(2018){Gaia Collaboration}, {Brown},
  {Vallenari}, {Prusti}, {de Bruijne}, {Babusiaux}, {Bailer-Jones}, {Biermann},
  {Evans}, {Eyer}, {Jansen}, {Jordi}, {Klioner}, {Lammers}, {Lindegren},
  {Luri}, {Mignard}, {Panem}, {Pourbaix}, {Randich}, {Sartoretti}, {Siddiqui},
  {Soubiran}, {van Leeuwen}, {Walton}, {Arenou}, {Bastian}, {Cropper},
  {Drimmel}, {Katz}, {Lattanzi}, {Bakker}, {Cacciari}, {Casta{\~n}eda},
  {Chaoul}, {Cheek}, {De Angeli}, {Fabricius}, {Guerra}, {Holl}, {Masana},
  {Messineo}, {Mowlavi}, {Nienartowicz}, {Panuzzo}, {Portell}, {Riello},
  {Seabroke}, {Tanga}, {Th{\'e}venin}, {Gracia-Abril}, {Comoretto},
  {Garcia-Reinaldos}, {Teyssier}, {Altmann}, {Andrae}, {Audard},
  {Bellas-Velidis}, {Benson}, {Berthier}, {Blomme}, {Burgess}, {Busso},
  {Carry}, {Cellino}, {Clementini}, {Clotet}, {Creevey}, {Davidson}, {De
  Ridder}, {Delchambre}, {Dell'Oro}, {Ducourant},
  {Fern{\'a}ndez-Hern{\'a}ndez}, {Fouesneau}, {Fr{\'e}mat}, {Galluccio},
  {Garc{\'\i}a-Torres}, {Gonz{\'a}lez-N{\'u}{\~n}ez}, {Gonz{\'a}lez-Vidal},
  {Gosset}, {Guy}, {Halbwachs}, {Hambly}, {Harrison}, {Hern{\'a}ndez},
  {Hestroffer}, {Hodgkin}, {Hutton}, {Jasniewicz}, {Jean-Antoine-Piccolo},
  {Jordan}, {Korn}, {Krone-Martins}, {Lanzafame}, {Lebzelter}, {L{\"o}ffler},
  {Manteiga}, {Marrese}, {Mart{\'\i}n-Fleitas}, {Moitinho}, {Mora}, {Muinonen},
  {Osinde}, {Pancino}, {Pauwels}, {Petit}, {Recio-Blanco}, {Richards},
  {Rimoldini}, {Robin}, {Sarro}, {Siopis}, {Smith}, {Sozzetti}, {S{\"u}veges},
  {Torra}, {van Reeven}, {Abbas}, {Abreu Aramburu}, {Accart}, {Aerts},
  {Altavilla}, {{\'A}lvarez}, {Alvarez}, {Alves}, {Anderson}, {Andrei},
  {Anglada Varela}, {Antiche}, {Antoja}, {Arcay}, {Astraatmadja}, {Bach},
  {Baker}, {Balaguer-N{\'u}{\~n}ez}, {Balm}, {Barache}, {Barata}, {Barbato},
  {Barblan}, {Barklem}, {Barrado}, {Barros}, {Barstow}, {Bartholom{\'e}
  Mu{\~n}oz}, {Bassilana}, {Becciani}, {Bellazzini}, {Berihuete}, {Bertone},
  {Bianchi}, {Bienaym{\'e}}, {Blanco-Cuaresma}, {Boch}, {Boeche}, {Bombrun},
  {Borrachero}, {Bossini}, {Bouquillon}, {Bourda}, {Bragaglia}, {Bramante},
  {Breddels}, {Bressan}, {Brouillet}, {Br{\"u}semeister}, {Brugaletta},
  {Bucciarelli}, {Burlacu}, {Busonero}, {Butkevich}, {Buzzi}, {Caffau},
  {Cancelliere}, {Cannizzaro}, {Cantat-Gaudin}, {Carballo}, {Carlucci},
  {Carrasco}, {Casamiquela}, {Castellani}, {Castro-Ginard}, {Charlot},
  {Chemin}, {Chiavassa}, {Cocozza}, {Costigan}, {Cowell}, {Crifo}, {Crosta},
  {Crowley}, {Cuypers}, {Dafonte}, {Damerdji}, {Dapergolas}, {David}, {David},
  {de Laverny}, {De Luise}, {De March}, {de Martino}, {de Souza}, {de Torres},
  {Debosscher}, {del Pozo}, {Delbo}, {Delgado}, {Delgado}, {Di Matteo},
  {Diakite}, {Diener}, {Distefano}, {Dolding}, {Drazinos}, {Dur{\'a}n},
  {Edvardsson}, {Enke}, {Eriksson}, {Esquej}, {Eynard Bontemps}, {Fabre},
  {Fabrizio}, {Faigler}, {Falc{\~a}o}, {Farr{\`a}s Casas}, {Federici},
  {Fedorets}, {Fernique}, {Figueras}, {Filippi}, {Findeisen}, {Fonti},
  {Fraile}, {Fraser}, {Fr{\'e}zouls}, {Gai}, {Galleti}, {Garabato},
  {Garc{\'\i}a-Sedano}, {Garofalo}, {Garralda}, {Gavel}, {Gavras}, {Gerssen},
  {Geyer}, {Giacobbe}, {Gilmore}, {Girona}, {Giuffrida}, {Glass}, {Gomes},
  {Granvik}, {Gueguen}, {Guerrier}, {Guiraud}, {Guti{\'e}rrez-S{\'a}nchez},
  {Haigron}, {Hatzidimitriou}, {Hauser}, {Haywood}, {Heiter}, {Helmi}, {Heu},
  {Hilger}, {Hobbs}, {Hofmann}, {Holland}, {Huckle}, {Hypki}, {Icardi},
  {Jan{\ss}en}, {Jevardat de Fombelle}, {Jonker}, {Juh{\'a}sz}, {Julbe},
  {Karampelas}, {Kewley}, {Klar}, {Kochoska}, {Kohley}, {Kolenberg},
  {Kontizas}, {Kontizas}, {Koposov}, {Kordopatis}, {Kostrzewa-Rutkowska},
  {Koubsky}, {Lambert}, {Lanza}, {Lasne}, {Lavigne}, {Le Fustec}, {Le
  Poncin-Lafitte}, {Lebreton}, {Leccia}, {Leclerc}, {Lecoeur-Taibi},
  {Lenhardt}, {Leroux}, {Liao}, {Licata}, {Lindstr{\o}m}, {Lister}, {Livanou},
  {Lobel}, {L{\'o}pez}, {Managau}, {Mann}, {Mantelet}, {Marchal}, {Marchant},
  {Marconi}, {Marinoni}, {Marschalk{\'o}}, {Marshall}, {Martino}, {Marton},
  {Mary}, {Massari}, {Matijevi{\v{c}}}, {Mazeh}, {McMillan}, {Messina},
  {Michalik}, {Millar}, {Molina}, {Molinaro}, {Moln{\'a}r}, {Montegriffo},
  {Mor}, {Morbidelli}, {Morel}, {Morris}, {Mulone}, {Muraveva}, {Musella},
  {Nelemans}, {Nicastro}, {Noval}, {O'Mullane}, {Ord{\'e}novic},
  {Ord{\'o}{\~n}ez-Blanco}, {Osborne}, {Pagani}, {Pagano}, {Pailler},
  {Palacin}, {Palaversa}, {Panahi}, {Pawlak}, {Piersimoni}, {Pineau}, {Plachy},
  {Plum}, {Poggio}, {Poujoulet}, {Pr{\v{s}}a}, {Pulone}, {Racero}, {Ragaini},
  {Rambaux}, {Ramos-Lerate}, {Regibo}, {Reyl{\'e}}, {Riclet}, {Ripepi}, {Riva},
  {Rivard}, {Rixon}, {Roegiers}, {Roelens}, {Romero-G{\'o}mez}, {Rowell},
  {Royer}, {Ruiz-Dern}, {Sadowski}, {Sagrist{\`a} Sell{\'e}s}, {Sahlmann},
  {Salgado}, {Salguero}, {Sanna}, {Santana-Ros}, {Sarasso}, {Savietto},
  {Schultheis}, {Sciacca}, {Segol}, {Segovia}, {S{\'e}gransan}, {Shih},
  {Siltala}, {Silva}, {Smart}, {Smith}, {Solano}, {Solitro}, {Sordo}, {Soria
  Nieto}, {Souchay}, {Spagna}, {Spoto}, {Stampa}, {Steele},
  {Steidelm{\"u}ller}, {Stephenson}, {Stoev}, {Suess}, {Surdej}, {Szabados},
  {Szegedi-Elek}, {Tapiador}, {Taris}, {Tauran}, {Taylor}, {Teixeira},
  {Terrett}, {Teyssand ier}, {Thuillot}, {Titarenko}, {Torra Clotet}, {Turon},
  {Ulla}, {Utrilla}, {Uzzi}, {Vaillant}, {Valentini}, {Valette}, {van Elteren},
  {Van Hemelryck}, {van Leeuwen}, {Vaschetto}, {Vecchiato}, {Veljanoski},
  {Viala}, {Vicente}, {Vogt}, {von Essen}, {Voss}, {Votruba}, {Voutsinas},
  {Walmsley}, {Weiler}, {Wertz}, {Wevers}, {Wyrzykowski}, {Yoldas},
  {{\v{Z}}erjal}, {Ziaeepour}, {Zorec}, {Zschocke}, {Zucker}, {Zurbach}, \&
  {Zwitter}}]{GaiaDR2}
{Gaia Collaboration}, {Brown}, A.~G.~A., {Vallenari}, A., {et~al.} 2018, \aap,
  616, A1

\bibitem[{{Gaia Collaboration} {et~al.}(2021){Gaia Collaboration}, {Brown},
  {Vallenari}, {Prusti}, {de Bruijne}, {Babusiaux}, {Biermann}, {Creevey},
  {Evans}, {Eyer}, {Hutton}, {Jansen}, {Jordi}, {Klioner}, {Lammers},
  {Lindegren}, {Luri}, {Mignard}, {Panem}, {Pourbaix}, {Randich}, {Sartoretti},
  {Soubiran}, {Walton}, {Arenou}, {Bailer-Jones}, {Bastian}, {Cropper},
  {Drimmel}, {Katz}, {Lattanzi}, {van Leeuwen}, {Bakker}, {Cacciari},
  {Casta{\~n}eda}, {De Angeli}, {Ducourant}, {Fabricius}, {Fouesneau},
  {Fr{\'e}mat}, {Guerra}, {Guerrier}, {Guiraud}, {Jean-Antoine Piccolo},
  {Masana}, {Messineo}, {Mowlavi}, {Nicolas}, {Nienartowicz}, {Pailler},
  {Panuzzo}, {Riclet}, {Roux}, {Seabroke}, {Sordo}, {Tanga}, {Th{\'e}venin},
  {Gracia-Abril}, {Portell}, {Teyssier}, {Altmann}, {Andrae}, {Bellas-Velidis},
  {Benson}, {Berthier}, {Blomme}, {Brugaletta}, {Burgess}, {Busso}, {Carry},
  {Cellino}, {Cheek}, {Clementini}, {Damerdji}, {Davidson}, {Delchambre},
  {Dell'Oro}, {Fern{\'a}ndez-Hern{\'a}ndez}, {Galluccio}, {Garc{\'\i}a-Lario},
  {Garcia-Reinaldos}, {Gonz{\'a}lez-N{\'u}{\~n}ez}, {Gosset}, {Haigron},
  {Halbwachs}, {Hambly}, {Harrison}, {Hatzidimitriou}, {Heiter},
  {Hern{\'a}ndez}, {Hestroffer}, {Hodgkin}, {Holl}, {Jan{\ss}en}, {Jevardat de
  Fombelle}, {Jordan}, {Krone-Martins}, {Lanzafame}, {L{\"o}ffler}, {Lorca},
  {Manteiga}, {Marchal}, {Marrese}, {Moitinho}, {Mora}, {Muinonen}, {Osborne},
  {Pancino}, {Pauwels}, {Petit}, {Recio-Blanco}, {Richards}, {Riello},
  {Rimoldini}, {Robin}, {Roegiers}, {Rybizki}, {Sarro}, {Siopis}, {Smith},
  {Sozzetti}, {Ulla}, {Utrilla}, {van Leeuwen}, {van Reeven}, {Abbas}, {Abreu
  Aramburu}, {Accart}, {Aerts}, {Aguado}, {Ajaj}, {Altavilla}, {{\'A}lvarez},
  {{\'A}lvarez Cid-Fuentes}, {Alves}, {Anderson}, {Anglada Varela}, {Antoja},
  {Audard}, {Baines}, {Baker}, {Balaguer-N{\'u}{\~n}ez}, {Balbinot}, {Balog},
  {Barache}, {Barbato}, {Barros}, {Barstow}, {Bartolom{\'e}}, {Bassilana},
  {Bauchet}, {Baudesson-Stella}, {Becciani}, {Bellazzini}, {Bernet}, {Bertone},
  {Bianchi}, {Blanco-Cuaresma}, {Boch}, {Bombrun}, {Bossini}, {Bouquillon},
  {Bragaglia}, {Bramante}, {Breedt}, {Bressan}, {Brouillet}, {Bucciarelli},
  {Burlacu}, {Busonero}, {Butkevich}, {Buzzi}, {Caffau}, {Cancelliere},
  {C{\'a}novas}, {Cantat-Gaudin}, {Carballo}, {Carlucci}, {Carnerero},
  {Carrasco}, {Casamiquela}, {Castellani}, {Castro-Ginard}, {Castro Sampol},
  {Chaoul}, {Charlot}, {Chemin}, {Chiavassa}, {Cioni}, {Comoretto}, {Cooper},
  {Cornez}, {Cowell}, {Crifo}, {Crosta}, {Crowley}, {Dafonte}, {Dapergolas},
  {David}, {David}, {de Laverny}, {De Luise}, {De March}, {De Ridder}, {de
  Souza}, {de Teodoro}, {de Torres}, {del Peloso}, {del Pozo}, {Delbo},
  {Delgado}, {Delgado}, {Delisle}, {Di Matteo}, {Diakite}, {Diener},
  {Distefano}, {Dolding}, {Eappachen}, {Edvardsson}, {Enke}, {Esquej}, {Fabre},
  {Fabrizio}, {Faigler}, {Fedorets}, {Fernique}, {Fienga}, {Figueras},
  {Fouron}, {Fragkoudi}, {Fraile}, {Franke}, {Gai}, {Garabato},
  {Garcia-Gutierrez}, {Garc{\'\i}a-Torres}, {Garofalo}, {Gavras}, {Gerlach},
  {Geyer}, {Giacobbe}, {Gilmore}, {Girona}, {Giuffrida}, {Gomel}, {Gomez},
  {Gonzalez-Santamaria}, {Gonz{\'a}lez-Vidal}, {Granvik},
  {Guti{\'e}rrez-S{\'a}nchez}, {Guy}, {Hauser}, {Haywood}, {Helmi}, {Hidalgo},
  {Hilger}, {H{\l}adczuk}, {Hobbs}, {Holland}, {Huckle}, {Jasniewicz},
  {Jonker}, {Juaristi Campillo}, {Julbe}, {Karbevska}, {Kervella}, {Khanna},
  {Kochoska}, {Kontizas}, {Kordopatis}, {Korn}, {Kostrzewa-Rutkowska},
  {Kruszy{\'n}ska}, {Lambert}, {Lanza}, {Lasne}, {Le Campion}, {Le Fustec},
  {Lebreton}, {Lebzelter}, {Leccia}, {Leclerc}, {Lecoeur-Taibi}, {Liao},
  {Licata}, {Lindstr{\o}m}, {Lister}, {Livanou}, {Lobel}, {Madrero Pardo},
  {Managau}, {Mann}, {Marchant}, {Marconi}, {Marcos Santos}, {Marinoni},
  {Marocco}, {Marshall}, {Martin Polo}, {Mart{\'\i}n-Fleitas}, {Masip},
  {Massari}, {Mastrobuono-Battisti}, {Mazeh}, {McMillan}, {Messina},
  {Michalik}, {Millar}, {Mints}, {Molina}, {Molinaro}, {Moln{\'a}r},
  {Montegriffo}, {Mor}, {Morbidelli}, {Morel}, {Morris}, {Mulone}, {Munoz},
  {Muraveva}, {Murphy}, {Musella}, {Noval}, {Ord{\'e}novic}, {Orr{\`u}},
  {Osinde}, {Pagani}, {Pagano}, {Palaversa}, {Palicio}, {Panahi}, {Pawlak},
  {Pe{\~n}alosa Esteller}, {Penttil{\"a}}, {Piersimoni}, {Pineau}, {Plachy},
  {Plum}, {Poggio}, {Poretti}, {Poujoulet}, {Pr{\v{s}}a}, {Pulone}, {Racero},
  {Ragaini}, {Rainer}, {Raiteri}, {Rambaux}, {Ramos}, {Ramos-Lerate}, {Re
  Fiorentin}, {Regibo}, {Reyl{\'e}}, {Ripepi}, {Riva}, {Rixon}, {Robichon},
  {Robin}, {Roelens}, {Rohrbasser}, {Romero-G{\'o}mez}, {Rowell}, {Royer},
  {Rybicki}, {Sadowski}, {Sagrist{\`a} Sell{\'e}s}, {Sahlmann}, {Salgado},
  {Salguero}, {Samaras}, {Sanchez Gimenez}, {Sanna}, {Santove{\~n}a},
  {Sarasso}, {Schultheis}, {Sciacca}, {Segol}, {Segovia}, {S{\'e}gransan},
  {Semeux}, {Shahaf}, {Siddiqui}, {Siebert}, {Siltala}, {Slezak}, {Smart},
  {Solano}, {Solitro}, {Souami}, {Souchay}, {Spagna}, {Spoto}, {Steele},
  {Steidelm{\"u}ller}, {Stephenson}, {S{\"u}veges}, {Szabados}, {Szegedi-Elek},
  {Taris}, {Tauran}, {Taylor}, {Teixeira}, {Thuillot}, {Tonello}, {Torra},
  {Torra}, {Turon}, {Unger}, {Vaillant}, {van Dillen}, {Vanel}, {Vecchiato},
  {Viala}, {Vicente}, {Voutsinas}, {Weiler}, {Wevers}, {Wyrzykowski}, {Yoldas},
  {Yvard}, {Zhao}, {Zorec}, {Zucker}, {Zurbach}, \& {Zwitter}}]{2020EDR3}
{Gaia Collaboration}, {Brown}, A.~G.~A., {Vallenari}, A., {et~al.} 2021, \aap,
  650, C3

\bibitem[{{Gaia Collaboration} {et~al.}(2016){Gaia Collaboration}, {Prusti},
  {de Bruijne}, {Brown}, {Vallenari}, {Babusiaux}, {Bailer-Jones}, {Bastian},
  {Biermann}, {Evans}, {Eyer}, {Jansen}, {Jordi}, {Klioner}, {Lammers},
  {Lindegren}, {Luri}, {Mignard}, {Milligan}, {Panem}, {Poinsignon},
  {Pourbaix}, {Randich}, {Sarri}, {Sartoretti}, {Siddiqui}, {Soubiran},
  {Valette}, {van Leeuwen}, {Walton}, {Aerts}, {Arenou}, {Cropper}, {Drimmel},
  {H{\o}g}, {Katz}, {Lattanzi}, {O'Mullane}, {Grebel}, {Holland}, {Huc},
  {Passot}, {Bramante}, {Cacciari}, {Casta{\~n}eda}, {Chaoul}, {Cheek}, {De
  Angeli}, {Fabricius}, {Guerra}, {Hern{\'a}ndez}, {Jean-Antoine-Piccolo},
  {Masana}, {Messineo}, {Mowlavi}, {Nienartowicz}, {Ord{\'o}{\~n}ez-Blanco},
  {Panuzzo}, {Portell}, {Richards}, {Riello}, {Seabroke}, {Tanga},
  {Th{\'e}venin}, {Torra}, {Els}, {Gracia-Abril}, {Comoretto},
  {Garcia-Reinaldos}, {Lock}, {Mercier}, {Altmann}, {Andrae}, {Astraatmadja},
  {Bellas-Velidis}, {Benson}, {Berthier}, {Blomme}, {Busso}, {Carry},
  {Cellino}, {Clementini}, {Cowell}, {Creevey}, {Cuypers}, {Davidson}, {De
  Ridder}, {de Torres}, {Delchambre}, {Dell'Oro}, {Ducourant}, {Fr{\'e}mat},
  {Garc{\'\i}a-Torres}, {Gosset}, {Halbwachs}, {Hambly}, {Harrison}, {Hauser},
  {Hestroffer}, {Hodgkin}, {Huckle}, {Hutton}, {Jasniewicz}, {Jordan},
  {Kontizas}, {Korn}, {Lanzafame}, {Manteiga}, {Moitinho}, {Muinonen},
  {Osinde}, {Pancino}, {Pauwels}, {Petit}, {Recio-Blanco}, {Robin}, {Sarro},
  {Siopis}, {Smith}, {Smith}, {Sozzetti}, {Thuillot}, {van Reeven}, {Viala},
  {Abbas}, {Abreu Aramburu}, {Accart}, {Aguado}, {Allan}, {Allasia},
  {Altavilla}, {{\'A}lvarez}, {Alves}, {Anderson}, {Andrei}, {Anglada Varela},
  {Antiche}, {Antoja}, {Ant{\'o}n}, {Arcay}, {Atzei}, {Ayache}, {Bach},
  {Baker}, {Balaguer-N{\'u}{\~n}ez}, {Barache}, {Barata}, {Barbier}, {Barblan},
  {Baroni}, {Barrado y Navascu{\'e}s}, {Barros}, {Barstow}, {Becciani},
  {Bellazzini}, {Bellei}, {Bello Garc{\'\i}a}, {Belokurov}, {Bendjoya},
  {Berihuete}, {Bianchi}, {Bienaym{\'e}}, {Billebaud}, {Blagorodnova},
  {Blanco-Cuaresma}, {Boch}, {Bombrun}, {Borrachero}, {Bouquillon}, {Bourda},
  {Bouy}, {Bragaglia}, {Breddels}, {Brouillet}, {Br{\"u}semeister},
  {Bucciarelli}, {Budnik}, {Burgess}, {Burgon}, {Burlacu}, {Busonero}, {Buzzi},
  {Caffau}, {Cambras}, {Campbell}, {Cancelliere}, {Cantat-Gaudin}, {Carlucci},
  {Carrasco}, {Castellani}, {Charlot}, {Charnas}, {Charvet}, {Chassat},
  {Chiavassa}, {Clotet}, {Cocozza}, {Collins}, {Collins}, {Costigan}, {Crifo},
  {Cross}, {Crosta}, {Crowley}, {Dafonte}, {Damerdji}, {Dapergolas}, {David},
  {David}, {De Cat}, {de Felice}, {de Laverny}, {De Luise}, {De March}, {de
  Martino}, {de Souza}, {Debosscher}, {del Pozo}, {Delbo}, {Delgado},
  {Delgado}, {di Marco}, {Di Matteo}, {Diakite}, {Distefano}, {Dolding}, {Dos
  Anjos}, {Drazinos}, {Dur{\'a}n}, {Dzigan}, {Ecale}, {Edvardsson}, {Enke},
  {Erdmann}, {Escolar}, {Espina}, {Evans}, {Eynard Bontemps}, {Fabre},
  {Fabrizio}, {Faigler}, {Falc{\~a}o}, {Farr{\`a}s Casas}, {Faye}, {Federici},
  {Fedorets}, {Fern{\'a}ndez-Hern{\'a}ndez}, {Fernique}, {Fienga}, {Figueras},
  {Filippi}, {Findeisen}, {Fonti}, {Fouesneau}, {Fraile}, {Fraser}, {Fuchs},
  {Furnell}, {Gai}, {Galleti}, {Galluccio}, {Garabato}, {Garc{\'\i}a-Sedano},
  {Gar{\'e}}, {Garofalo}, {Garralda}, {Gavras}, {Gerssen}, {Geyer}, {Gilmore},
  {Girona}, {Giuffrida}, {Gomes}, {Gonz{\'a}lez-Marcos},
  {Gonz{\'a}lez-N{\'u}{\~n}ez}, {Gonz{\'a}lez-Vidal}, {Granvik}, {Guerrier},
  {Guillout}, {Guiraud}, {G{\'u}rpide}, {Guti{\'e}rrez-S{\'a}nchez}, {Guy},
  {Haigron}, {Hatzidimitriou}, {Haywood}, {Heiter}, {Helmi}, {Hobbs},
  {Hofmann}, {Holl}, {Holland}, {Hunt}, {Hypki}, {Icardi}, {Irwin}, {Jevardat
  de Fombelle}, {Jofr{\'e}}, {Jonker}, {Jorissen}, {Julbe}, {Karampelas},
  {Kochoska}, {Kohley}, {Kolenberg}, {Kontizas}, {Koposov}, {Kordopatis},
  {Koubsky}, {Kowalczyk}, {Krone-Martins}, {Kudryashova}, {Kull}, {Bachchan},
  {Lacoste-Seris}, {Lanza}, {Lavigne}, {Le Poncin-Lafitte}, {Lebreton},
  {Lebzelter}, {Leccia}, {Leclerc}, {Lecoeur-Taibi}, {Lemaitre}, {Lenhardt},
  {Leroux}, {Liao}, {Licata}, {Lindstr{\o}m}, {Lister}, {Livanou}, {Lobel},
  {L{\"o}ffler}, {L{\'o}pez}, {Lopez-Lozano}, {Lorenz}, {Loureiro},
  {MacDonald}, {Magalh{\~a}es Fernandes}, {Managau}, {Mann}, {Mantelet},
  {Marchal}, {Marchant}, {Marconi}, {Marie}, {Marinoni}, {Marrese},
  {Marschalk{\'o}}, {Marshall}, {Mart{\'\i}n-Fleitas}, {Martino}, {Mary},
  {Matijevi{\v{c}}}, {Mazeh}, {McMillan}, {Messina}, {Mestre}, {Michalik},
  {Millar}, {Miranda}, {Molina}, {Molinaro}, {Molinaro}, {Moln{\'a}r},
  {Moniez}, {Montegriffo}, {Monteiro}, {Mor}, {Mora}, {Morbidelli}, {Morel},
  {Morgenthaler}, {Morley}, {Morris}, {Mulone}, {Muraveva}, {Musella},
  {Narbonne}, {Nelemans}, {Nicastro}, {Noval}, {Ord{\'e}novic},
  {Ordieres-Mer{\'e}}, {Osborne}, {Pagani}, {Pagano}, {Pailler}, {Palacin},
  {Palaversa}, {Parsons}, {Paulsen}, {Pecoraro}, {Pedrosa}, {Pentik{\"a}inen},
  {Pereira}, {Pichon}, {Piersimoni}, {Pineau}, {Plachy}, {Plum}, {Poujoulet},
  {Pr{\v{s}}a}, {Pulone}, {Ragaini}, {Rago}, {Rambaux}, {Ramos-Lerate},
  {Ranalli}, {Rauw}, {Read}, {Regibo}, {Renk}, {Reyl{\'e}}, {Ribeiro},
  {Rimoldini}, {Ripepi}, {Riva}, {Rixon}, {Roelens}, {Romero-G{\'o}mez},
  {Rowell}, {Royer}, {Rudolph}, {Ruiz-Dern}, {Sadowski}, {Sagrist{\`a}
  Sell{\'e}s}, {Sahlmann}, {Salgado}, {Salguero}, {Sarasso}, {Savietto},
  {Schnorhk}, {Schultheis}, {Sciacca}, {Segol}, {Segovia}, {Segransan},
  {Serpell}, {Shih}, {Smareglia}, {Smart}, {Smith}, {Solano}, {Solitro},
  {Sordo}, {Soria Nieto}, {Souchay}, {Spagna}, {Spoto}, {Stampa}, {Steele},
  {Steidelm{\"u}ller}, {Stephenson}, {Stoev}, {Suess}, {S{\"u}veges}, {Surdej},
  {Szabados}, {Szegedi-Elek}, {Tapiador}, {Taris}, {Tauran}, {Taylor},
  {Teixeira}, {Terrett}, {Tingley}, {Trager}, {Turon}, {Ulla}, {Utrilla},
  {Valentini}, {van Elteren}, {Van Hemelryck}, {van Leeuwen}, {Varadi},
  {Vecchiato}, {Veljanoski}, {Via}, {Vicente}, {Vogt}, {Voss}, {Votruba},
  {Voutsinas}, {Walmsley}, {Weiler}, {Weingrill}, {Werner}, {Wevers},
  {Whitehead}, {Wyrzykowski}, {Yoldas}, {{\v{Z}}erjal}, {Zucker}, {Zurbach},
  {Zwitter}, {Alecu}, {Allen}, {Allende Prieto}, {Amorim},
  {Anglada-Escud{\'e}}, {Arsenijevic}, {Azaz}, {Balm}, {Beck}, {Bernstein},
  {Bigot}, {Bijaoui}, {Blasco}, {Bonfigli}, {Bono}, {Boudreault}, {Bressan},
  {Brown}, {Brunet}, {Bunclark}, {Buonanno}, {Butkevich}, {Carret}, {Carrion},
  {Chemin}, {Ch{\'e}reau}, {Corcione}, {Darmigny}, {de Boer}, {de Teodoro}, {de
  Zeeuw}, {Delle Luche}, {Domingues}, {Dubath}, {Fodor}, {Fr{\'e}zouls},
  {Fries}, {Fustes}, {Fyfe}, {Gallardo}, {Gallegos}, {Gardiol}, {Gebran},
  {Gomboc}, {G{\'o}mez}, {Grux}, {Gueguen}, {Heyrovsky}, {Hoar}, {Iannicola},
  {Isasi Parache}, {Janotto}, {Joliet}, {Jonckheere}, {Keil}, {Kim},
  {Klagyivik}, {Klar}, {Knude}, {Kochukhov}, {Kolka}, {Kos}, {Kutka}, {Lainey},
  {LeBouquin}, {Liu}, {Loreggia}, {Makarov}, {Marseille}, {Martayan},
  {Martinez-Rubi}, {Massart}, {Meynadier}, {Mignot}, {Munari}, {Nguyen},
  {Nordlander}, {Ocvirk}, {O'Flaherty}, {Olias Sanz}, {Ortiz}, {Osorio},
  {Oszkiewicz}, {Ouzounis}, {Palmer}, {Park}, {Pasquato}, {Peltzer}, {Peralta},
  {P{\'e}turaud}, {Pieniluoma}, {Pigozzi}, {Poels}, {Prat}, {Prod'homme},
  {Raison}, {Rebordao}, {Risquez}, {Rocca-Volmerange}, {Rosen}, {Ruiz-Fuertes},
  {Russo}, {Sembay}, {Serraller Vizcaino}, {Short}, {Siebert}, {Silva},
  {Sinachopoulos}, {Slezak}, {Soffel}, {Sosnowska}, {Strai{\v{z}}ys}, {ter
  Linden}, {Terrell}, {Theil}, {Tiede}, {Troisi}, {Tsalmantza}, {Tur},
  {Vaccari}, {Vachier}, {Valles}, {Van Hamme}, {Veltz}, {Virtanen}, {Wallut},
  {Wichmann}, {Wilkinson}, {Ziaeepour}, \& {Zschocke}}]{2016GaiaMission}
{Gaia Collaboration}, {Prusti}, T., {de Bruijne}, J.~H.~J., {et~al.} 2016,
  \aap, 595, A1

\bibitem[{{Gould}(1992)}]{1992Gould}
{Gould}, A. 1992, \apj, 392, 442

\bibitem[{{Gould}(2000)}]{2000ApJ...542..785G}
{Gould}, A. 2000, \apj, 542, 785

\bibitem[{{Gould}(2004)}]{2004Gould}
{Gould}, A. 2004, \apj, 606, 319

\bibitem[{{Gould} {et~al.}(2010){Gould}, {Dong}, {Gaudi}, {Udalski}, {Bond},
  {Greenhill}, {Street}, {Dominik}, {Sumi}, {Szyma{\'n}ski}, {Han}, {Allen},
  {Bolt}, {Bos}, {Christie}, {DePoy}, {Drummond}, {Eastman}, {Gal-Yam},
  {Higgins}, {Janczak}, {Kaspi}, {Koz{\l}owski}, {Lee}, {Mallia}, {Maury},
  {Maoz}, {McCormick}, {Monard}, {Moorhouse}, {Morgan}, {Natusch}, {Ofek},
  {Park}, {Pogge}, {Polishook}, {Santallo}, {Shporer}, {Spector}, {Thornley},
  {Yee}, {{\ensuremath{\mu}}FUN Collaboration}, {Kubiak}, {Pietrzy{\'n}ski},
  {Soszy{\'n}ski}, {Szewczyk}, {Wyrzykowski}, {Ulaczyk}, {Poleski}, {OGLE
  Collaboration}, {Abe}, {Bennett}, {Botzler}, {Douchin}, {Freeman}, {Fukui},
  {Furusawa}, {Hearnshaw}, {Hosaka}, {Itow}, {Kamiya}, {Kilmartin}, {Korpela},
  {Lin}, {Ling}, {Makita}, {Masuda}, {Matsubara}, {Miyake}, {Muraki}, {Nagaya},
  {Nishimoto}, {Ohnishi}, {Okumura}, {Perrott}, {Philpott}, {Rattenbury},
  {Saito}, {Sako}, {Sullivan}, {Sweatman}, {Tristram}, {von Seggern}, {Yock},
  {MOA Collaboration}, {Albrow}, {Batista}, {Beaulieu}, {Brillant}, {Caldwell},
  {Calitz}, {Cassan}, {Cole}, {Cook}, {Coutures}, {Dieters}, {Dominis Prester},
  {Donatowicz}, {Fouqu{\'e}}, {Hill}, {Hoffman}, {Jablonski}, {Kane}, {Kains},
  {Kubas}, {Marquette}, {Martin}, {Martioli}, {Meintjes}, {Menzies},
  {Pedretti}, {Pollard}, {Sahu}, {Vinter}, {Wambsganss}, {Watson}, {Williams},
  {Zub}, {PLANET Collaboration}, {Allan}, {Bode}, {Bramich}, {Burgdorf},
  {Clay}, {Fraser}, {Hawkins}, {Horne}, {Kerins}, {Lister}, {Mottram},
  {Saunders}, {Snodgrass}, {Steele}, {Tsapras}, {RoboNet Collaboration},
  {J{\o}rgensen}, {Anguita}, {Bozza}, {Calchi Novati}, {Harps{\o}e}, {Hinse},
  {Hundertmark}, {Kj{\ae}rgaard}, {Liebig}, {Mancini}, {Masi}, {Mathiasen},
  {Rahvar}, {Ricci}, {Scarpetta}, {Southworth}, {Surdej}, {Th{\"o}ne}, \&
  {MiNDSTEp Consortium}}]{2010microFun}
{Gould}, A., {Dong}, S., {Gaudi}, B.~S., {et~al.} 2010, \apj, 720, 1073

\bibitem[{{Gould} {et~al.}(2009){Gould}, {Udalski}, {Monard}, {Horne}, {Dong},
  {Miyake}, {Sahu}, {Bennett}, {Wyrzykowski}, {Soszy{\'n}ski}, {Szyma{\'n}ski},
  {Kubiak}, {Pietrzy{\'n}ski}, {Szewczyk}, {Ulaczyk}, {OGLE Collaboration},
  {Allen}, {Christie}, {DePoy}, {Gaudi}, {Han}, {Lee}, {McCormick}, {Natusch},
  {Park}, {Pogge}, {{\ensuremath{\mu}}FUN Collaboration}, {Allan}, {Bode},
  {Bramich}, {Burgdorf}, {Dominik}, {Fraser}, {Kerins}, {Mottram}, {Snodgrass},
  {Steele}, {Street}, {Tsapras}, {RoboNet Collaboration}, {Abe}, {Bond},
  {Botzler}, {Fukui}, {Furusawa}, {Hearnshaw}, {Itow}, {Kamiya}, {Kilmartin},
  {Korpela}, {Lin}, {Ling}, {Masuda}, {Matsubara}, {Muraki}, {Nagaya},
  {Ohnishi}, {Okumura}, {Perrott}, {Rattenbury}, {Saito}, {Sako}, {Skuljan},
  {Sullivan}, {Sumi}, {Sweatman}, {Tristram}, {Yock}, {MOA Collaboration},
  {Albrow}, {Beaulieu}, {Coutures}, {Calitz}, {Caldwell}, {Fouque}, {Martin},
  {Williams}, \& {PLANET Collaboration}}]{2009ApJ...698L.147G}
{Gould}, A., {Udalski}, A., {Monard}, B., {et~al.} 2009, \apjl, 698, L147

\bibitem[{{Hardy} \& {Walker}(1995)}]{1995MNRAS.276L..79H}
{Hardy}, S.~J. \& {Walker}, M.~A. 1995, \mnras, 276, L79

\bibitem[{{Hobbs} {et~al.}(2005){Hobbs}, {Lorimer}, {Lyne}, \&
  {Kramer}}]{2005HobbsNS}
{Hobbs}, G., {Lorimer}, D.~R., {Lyne}, A.~G., \& {Kramer}, M. 2005, \mnras,
  360, 974

\bibitem[{{Hodgkin} {et~al.}(2021){Hodgkin}, {Harrison}, {Breedt}, {Wevers},
  {Rixon}, {Delgado}, {Yoldas}, {Kostrzewa-Rutkowska}, {Wyrzykowski}, {van
  Leeuwen}, {Blagorodnova}, {Campbell}, {Eappachen}, {Fraser}, {Ihanec},
  {Koposov}, {Kruszy{\'n}ska}, {Marton}, {Rybicki}, {Brown}, {Burgess},
  {Busso}, {Cowell}, {De Angeli}, {Diener}, {Evans}, {Gilmore}, {Holland},
  {Jonker}, {van Leeuwen}, {Mignard}, {Osborne}, {Portell}, {Prusti},
  {Richards}, {Riello}, {Seabroke}, {Walton}, {{\'A}brah{\'a}m}, {Altavilla},
  {Baker}, {Bastian}, {O'Brien}, {de Bruijne}, {Butterley}, {Carrasco},
  {Casta{\~n}eda}, {Clark}, {Clementini}, {Copperwheat}, {Cropper},
  {Damljanovic}, {Davidson}, {Davis}, {Dennefeld}, {Dhillon}, {Dolding},
  {Dominik}, {Esquej}, {Eyer}, {Fabricius}, {Fridman}, {Froebrich}, {Garralda},
  {Gomboc}, {Gonz{\'a}lez-Vidal}, {Guerra}, {Hambly}, {Hardy}, {Holl},
  {Hourihane}, {Japelj}, {Kann}, {Kiss}, {Knigge}, {Kolb}, {Komossa},
  {K{\'o}sp{\'a}l}, {Kov{\'a}cs}, {Kun}, {Leto}, {Lewis}, {Littlefair},
  {Mahabal}, {Mundell}, {Nagy}, {Padeletti}, {Palaversa}, {Pigulski},
  {Pretorius}, {van Reeven}, {Ribeiro}, {Roelens}, {Rowell}, {Schartel},
  {Scholz}, {Schwope}, {Sip{\H{o}}cz}, {Smartt}, {Smith}, {Serraller},
  {Steeghs}, {Sullivan}, {Szabados}, {Szegedi-Elek}, {Tisserand}, {Tomasella},
  {van Velzen}, {Whitelock}, {Wilson}, \& {Young}}]{2021Hodgkin}
{Hodgkin}, S.~T., {Harrison}, D.~L., {Breedt}, E., {et~al.} 2021, \aap, 652,
  A76

\bibitem[{{Hodgkin} {et~al.}(2013){Hodgkin}, {Wyrzykowski}, {Blagorodnova}, \&
  {Koposov}}]{2013RSPTA.37120239H}
{Hodgkin}, S.~T., {Wyrzykowski}, L., {Blagorodnova}, N., \& {Koposov}, S. 2013,
  Philosophical Transactions of the Royal Society of London Series A, 371,
  20120239

\bibitem[{{Holz} \& {Wald}(1996)}]{1996ApJ...471...64H}
{Holz}, D.~E. \& {Wald}, R.~M. 1996, \apj, 471, 64

\bibitem[{{Jayasinghe} {et~al.}(2021){Jayasinghe}, {Stanek}, {Thompson},
  {Kochanek}, {Rowan}, {Vallely}, {Strassmeier}, {Weber}, {Hinkle}, {Hambsch},
  {Martin}, {Prieto}, {Pessi}, {Huber}, {Auchettl}, {Lopez}, {Ilyin},
  {Badenes}, {Howard}, {Isaacson}, \& {Murphy}}]{2021UnicornBH}
{Jayasinghe}, T., {Stanek}, K.~Z., {Thompson}, T.~A., {et~al.} 2021, \mnras,
  504, 2577

\bibitem[{{Jeong} {et~al.}(1999){Jeong}, {Han}, \& {Park}}]{1999Jeong}
{Jeong}, Y., {Han}, C., \& {Park}, S.-H. 1999, \apj, 511, 569

\bibitem[{{Jordi} {et~al.}(2010){Jordi}, {Gebran}, {Carrasco}, {de Bruijne},
  {Voss}, {Fabricius}, {Knude}, {Vallenari}, {Kohley}, \&
  {Mora}}]{2010JordiGaiaPhot}
{Jordi}, C., {Gebran}, M., {Carrasco}, J.~M., {et~al.} 2010, \aap, 523, A48

\bibitem[{{Karolinski} \& {Zhu}(2020)}]{2020KarolinskiZhu}
{Karolinski}, N. \& {Zhu}, W. 2020, \mnras, 498, L25

\bibitem[{{Kim} {et~al.}(2016){Kim}, {Lee}, {Park}, {Kim}, {Cha}, {Lee}, {Han},
  {Chun}, \& {Yuk}}]{2016JKAS...49...37K}
{Kim}, S.-L., {Lee}, C.-U., {Park}, B.-G., {et~al.} 2016, Journal of Korean
  Astronomical Society, 49, 37

\bibitem[{{Koshimoto} {et~al.}(2020){Koshimoto}, {Bennett}, \&
  {Suzuki}}]{2020Koshimoto}
{Koshimoto}, N., {Bennett}, D.~P., \& {Suzuki}, D. 2020, \aj, 159, 268

\bibitem[{{Koshimoto} {et~al.}(2017){Koshimoto}, {Shvartzvald}, {Bennett},
  {Penny}, {Hundertmark}, {Bond}, {Zang}, {Henderson}, {Suzuki}, {Rattenbury},
  {Sumi}, {and}, {Abe}, {Asakura}, {Bhattacharya}, {Donachie}, {Evans},
  {Fukui}, {Hirao}, {Itow}, {Li}, {Ling}, {Masuda}, {Matsubara}, {Matsuo},
  {Muraki}, {Nagakane}, {Ohnishi}, {Ranc}, {Saito}, {Sharan}, {Shibai},
  {Sullivan}, {Tristram}, {Yamada}, {Yamada}, {Yonehara}, {MOA Collaboration},
  {Gelino}, {Beichman}, {Beaulieu}, {Marquette}, {Batista}, {Keck Team},
  {Friedmann}, {Hallakoun}, {Kaspi}, {Maoz}, {Wise Group}, {Bryden}, {Calchi
  Novati}, {Howell}, {UKIRT Team}, {Wang}, {Mao}, {Fouqu{\'e}}, {Microlensing
  Survey}, {Korhonen}, {J{\o}rgensen}, {Street}, {Tsapras}, {Dominik},
  {Kerins}, {Cassan}, {Snodgrass}, {Bachelet}, {Bozza}, {Bramich}, \& {VST-K2C9
  Team}}]{2017Koshimoto}
{Koshimoto}, N., {Shvartzvald}, Y., {Bennett}, D.~P., {et~al.} 2017, \aj, 154,
  3

\bibitem[{{Kreidberg} {et~al.}(2012){Kreidberg}, {Bailyn}, {Farr}, \&
  {Kalogera}}]{2012KreidbergNomassgap}
{Kreidberg}, L., {Bailyn}, C.~D., {Farr}, W.~M., \& {Kalogera}, V. 2012, \apj,
  757, 36

\bibitem[{{Lallement} {et~al.}(2018){Lallement}, {Capitanio}, {Ruiz-Dern},
  {Danielski}, {Babusiaux}, {Vergely}, {Elyajouri}, {Arenou}, \&
  {Leclerc}}]{Reddening3}
{Lallement}, R., {Capitanio}, L., {Ruiz-Dern}, L., {et~al.} 2018, \aap, 616,
  A132

\bibitem[{{Lallement} {et~al.}(2014){Lallement}, {Vergely}, {Valette},
  {Puspitarini}, {Eyer}, \& {Casagrande}}]{Reddening}
{Lallement}, R., {Vergely}, J.~L., {Valette}, B., {et~al.} 2014, \aap, 561, A91

\bibitem[{{Lam} {et~al.}(2020){Lam}, {Lu}, {Hosek}, {Dawson}, \&
  {Golovich}}]{2020LamPopSyCLE}
{Lam}, C.~Y., {Lu}, J.~R., {Hosek}, Matthew~W., J., {Dawson}, W.~A., \&
  {Golovich}, N.~R. 2020, \apj, 889, 31

\bibitem[{{Liebes}(1964)}]{1964Liebes}
{Liebes}, S. 1964, Physical Review, 133, 835

\bibitem[{{Lindegren} {et~al.}(2021){Lindegren}, {Klioner}, {Hern{\'a}ndez},
  {Bombrun}, {Ramos-Lerate}, {Steidelm{\"u}ller}, {Bastian}, {Biermann}, {de
  Torres}, {Gerlach}, {Geyer}, {Hilger}, {Hobbs}, {Lammers}, {McMillan},
  {Stephenson}, {Casta{\~n}eda}, {Davidson}, {Fabricius}, {Gracia-Abril},
  {Portell}, {Rowell}, {Teyssier}, {Torra}, {Bartolom{\'e}}, {Clotet},
  {Garralda}, {Gonz{\'a}lez-Vidal}, {Torra}, {Abbas}, {Altmann}, {Anglada
  Varela}, {Balaguer-N{\'u}{\~n}ez}, {Balog}, {Barache}, {Becciani}, {Bernet},
  {Bertone}, {Bianchi}, {Bouquillon}, {Brown}, {Bucciarelli}, {Busonero},
  {Butkevich}, {Buzzi}, {Cancelliere}, {Carlucci}, {Charlot}, {Cioni},
  {Crosta}, {Crowley}, {del Peloso}, {del Pozo}, {Drimmel}, {Esquej}, {Fienga},
  {Fraile}, {Gai}, {Garcia-Reinaldos}, {Guerra}, {Hambly}, {Hauser},
  {Jan{\ss}en}, {Jordan}, {Kostrzewa-Rutkowska}, {Lattanzi}, {Liao}, {Licata},
  {Lister}, {L{\"o}ffler}, {Marchant}, {Masip}, {Mignard}, {Mints}, {Molina},
  {Mora}, {Morbidelli}, {Murphy}, {Pagani}, {Panuzzo}, {Pe{\~n}alosa Esteller},
  {Poggio}, {Re Fiorentin}, {Riva}, {Sagrist{\`a} Sell{\'e}s}, {Sanchez
  Gimenez}, {Sarasso}, {Sciacca}, {Siddiqui}, {Smart}, {Souami}, {Spagna},
  {Steele}, {Taris}, {Utrilla}, {van Reeven}, \&
  {Vecchiato}}]{2020EDR3Astrometry}
{Lindegren}, L., {Klioner}, S.~A., {Hern{\'a}ndez}, J., {et~al.} 2021, \aap,
  649, A2

\bibitem[{{Mao} \& {Paczynski}(1991)}]{1991ApJ...374L..37M}
{Mao}, S. \& {Paczynski}, B. 1991, \apjl, 374, L37

\bibitem[{{Mao} {et~al.}(2002){Mao}, {Smith}, {Wo{\'z}niak}, {Udalski},
  {Szyma{\'n}ski}, {Kubiak}, {Pietrzy{\'n}ski}, {Soszy{\'n}ski}, \&
  {{\.Z}ebru{\'n}}}]{2002MaoLongUlens}
{Mao}, S., {Smith}, M.~C., {Wo{\'z}niak}, P., {et~al.} 2002, \mnras, 329, 349

\bibitem[{{McMahon} {et~al.}(2013){McMahon}, {Banerji}, {Gonzalez}, {Koposov},
  {Bejar}, {Lodieu}, {Rebolo}, \& {VHS Collaboration}}]{2013McMahon}
{McMahon}, R.~G., {Banerji}, M., {Gonzalez}, E., {et~al.} 2013, The Messenger,
  154, 35

\bibitem[{{Mr{\'o}z} {et~al.}(2020){Mr{\'o}z}, {Udalski}, {Szyma{\'n}ski},
  {Soszy{\'n}ski}, {Pietrukowicz}, {Koz{\l}owski}, {Skowron}, {Poleski},
  {Ulaczyk}, {Gromadzki}, {Rybicki}, {Iwanek}, \& {Wrona}}]{2020MrozDisk}
{Mr{\'o}z}, P., {Udalski}, A., {Szyma{\'n}ski}, M.~K., {et~al.} 2020, \apjs,
  249, 16

\bibitem[{{Mr{\'o}z} \& {Wyrzykowski}(2021)}]{MW}
{Mr{\'o}z}, P. \& {Wyrzykowski}, {\L}. 2021, \actaa, 71, 89

\bibitem[{{{\"O}zel} {et~al.}(2010){{\"O}zel}, {Psaltis}, {Narayan}, \&
  {McClintock}}]{2010OzelBHmass}
{{\"O}zel}, F., {Psaltis}, D., {Narayan}, R., \& {McClintock}, J.~E. 2010,
  \apj, 725, 1918

\bibitem[{{Paczynski}(1986)}]{1986Paczynski}
{Paczynski}, B. 1986, \apj, 304, 1

\bibitem[{{Paczynski}(1996)}]{1996ARA&A..34..419P}
{Paczynski}, B. 1996, \araa, 34, 419

\bibitem[{{Pecaut} \& {Mamajek}(2013)}]{2013PecautMamajek}
{Pecaut}, M.~J. \& {Mamajek}, E.~E. 2013, \apjs, 208, 9

\bibitem[{{Poindexter} {et~al.}(2005){Poindexter}, {Afonso}, {Bennett},
  {Glicenstein}, {Gould}, {Szyma{\'n}ski}, \& {Udalski}}]{2005ApJ...633..914P}
{Poindexter}, S., {Afonso}, C., {Bennett}, D.~P., {et~al.} 2005, \apj, 633, 914

\bibitem[{{Poleski} \& {Yee}(2019)}]{MulensModel}
{Poleski}, R. \& {Yee}, J.~C. 2019, Astronomy and Computing, 26, 35

\bibitem[{{Refsdal}(1964)}]{1964Refsdal}
{Refsdal}, S. 1964, \mnras, 128, 295

\bibitem[{{Refsdal}(1966)}]{1966MNRAS.134..315R}
{Refsdal}, S. 1966, \mnras, 134, 315

\bibitem[{{Riello} {et~al.}(2021){Riello}, {De Angeli}, {Evans}, {Montegriffo},
  {Carrasco}, {Busso}, {Palaversa}, {Burgess}, {Diener}, {Davidson}, {Rowell},
  {Fabricius}, {Jordi}, {Bellazzini}, {Pancino}, {Harrison}, {Cacciari}, {van
  Leeuwen}, {Hambly}, {Hodgkin}, {Osborne}, {Altavilla}, {Barstow}, {Brown},
  {Castellani}, {Cowell}, {De Luise}, {Gilmore}, {Giuffrida}, {Hidalgo},
  {Holland}, {Marinoni}, {Pagani}, {Piersimoni}, {Pulone}, {Ragaini}, {Rainer},
  {Richards}, {Sanna}, {Walton}, {Weiler}, \& {Yoldas}}]{2021RielloPhoto}
{Riello}, M., {De Angeli}, F., {Evans}, D.~W., {et~al.} 2021, \aap, 649, A3

\bibitem[{{Rybicki} {et~al.}(2022){Rybicki}, {Wyrzykowski}, {Bachelet},
  {Cassan}, {Zieli{\'n}ski}, {Gould}, {Calchi Novati}, {Yee}, {Ryu},
  {Gromadzki}, {Miko{\l}ajczyk}, {Ihanec}, {Kruszy{\'n}ska}, {Hambsch},
  {Zo{\l}a}, {Fossey}, {Awiphan}, {Nakharutai}, {Lewis}, {Olivares E.},
  {Hodgkin}, {Delgado}, {Breedt}, {Harrison}, {van Leeuwen}, {Rixon}, {Wevers},
  {Yoldas}, {Udalski}, {Szyma{\'n}ski}, {Soszy{\'n}ski}, {Pietrukowicz},
  {Koz{\l}owski}, {Skowron}, {Poleski}, {Ulaczyk}, {Mr{\'o}z}, {Iwanek},
  {Wrona}, {Street}, {Tsapras}, {Hundertmark}, {Dominik}, {Beichman}, {Bryden},
  {Carey}, {Gaudi}, {Henderson}, {Shvartzvald}, {Zang}, {Zhu}, {Christie},
  {Green}, {Hennerley}, {McCormick}, {Monard}, {Natusch}, {Pogge}, {Gezer},
  {Gurgul}, {Kaczmarek}, {Konacki}, {Lam}, {Maskoliunas}, {Pakstiene},
  {Ratajczak}, {Stankeviciute}, {Zdanavicius}, \&
  {Zi{\'o}{\l}kowska}}]{RybickiGaia19bld}
{Rybicki}, K.~A., {Wyrzykowski}, {\L}., {Bachelet}, E., {et~al.} 2022, \aap,
  657, A18

\bibitem[{{Rybicki} {et~al.}(2018){Rybicki}, {Wyrzykowski}, {Klencki}, {de
  Bruijne}, {Belczy{\'n}ski}, \& {Chru{\'s}li{\'n}ska}}]{2018MNRAS.476.2013R}
{Rybicki}, K.~A., {Wyrzykowski}, {\L}., {Klencki}, J., {et~al.} 2018, \mnras,
  476, 2013

\bibitem[{{Sahu} {et~al.}(2017){Sahu}, {Anderson}, {Casertano}, {Bond},
  {Bergeron}, {Nelan}, {Pueyo}, {Brown}, {Bellini}, {Levay}, {Sokol}, {aff1},
  {Dominik}, {Calamida}, {Kains}, \& {Livio}}]{2017Sci...356.1046S}
{Sahu}, K.~C., {Anderson}, J., {Casertano}, S., {et~al.} 2017, Science, 356,
  1046

\bibitem[{{Schlafly} \& {Finkbeiner}(2011)}]{2011Schlafly}
{Schlafly}, E.~F. \& {Finkbeiner}, D.~P. 2011, \apj, 737, 103

\bibitem[{{Skowron} {et~al.}(2011){Skowron}, {Udalski}, {Gould}, {Dong},
  {Monard}, {Han}, {Nelson}, {McCormick}, {Moorhouse}, {Thornley}, {Maury},
  {Bramich}, {Greenhill}, {Koz{\l}owski}, {Bond}, {Poleski}, {Wyrzykowski},
  {Ulaczyk}, {Kubiak}, {Szyma{\'n}ski}, {Pietrzy{\'n}ski}, {Soszy{\'n}ski},
  {OGLE Collaboration}, {Gaudi}, {Yee}, {Hung}, {Pogge}, {DePoy}, {Lee},
  {Park}, {Allen}, {Mallia}, {Drummond}, {Bolt}, {{\ensuremath{\mu}}FUN
  Collaboration}, {Allan}, {Browne}, {Clay}, {Dominik}, {Fraser}, {Horne},
  {Kains}, {Mottram}, {Snodgrass}, {Steele}, {Street}, {Tsapras}, {RoboNet
  Collaboration}, {Abe}, {Bennett}, {Botzler}, {Douchin}, {Freeman}, {Fukui},
  {Furusawa}, {Hayashi}, {Hearnshaw}, {Hosaka}, {Itow}, {Kamiya}, {Kilmartin},
  {Korpela}, {Lin}, {Ling}, {Makita}, {Masuda}, {Matsubara}, {Muraki},
  {Nagayama}, {Miyake}, {Nishimoto}, {Ohnishi}, {Perrott}, {Rattenbury},
  {Saito}, {Skuljan}, {Sullivan}, {Sumi}, {Suzuki}, {Sweatman}, {Tristram},
  {Wada}, {Yock}, {MOA Collaboration}, {Beaulieu}, {Fouqu{\'e}}, {Albrow},
  {Batista}, {Brillant}, {Caldwell}, {Cassan}, {Cole}, {Cook}, {Coutures},
  {Dieters}, {Dominis Prester}, {Donatowicz}, {Kane}, {Kubas}, {Marquette},
  {Martin}, {Menzies}, {Sahu}, {Wambsganss}, {Williams}, {Zub}, \& {PLANET
  Collaboration}}]{2011Skowron}
{Skowron}, J., {Udalski}, A., {Gould}, A., {et~al.} 2011, \apj, 738, 87

\bibitem[{{Stassun} {et~al.}(2019){Stassun}, {Oelkers}, {Paegert}, {Torres},
  {Pepper}, {De Lee}, {Collins}, {Latham}, {Muirhead}, {Chittidi},
  {Rojas-Ayala}, {Fleming}, {Rose}, {Tenenbaum}, {Ting}, {Kane}, {Barclay},
  {Bean}, {Brassuer}, {Charbonneau}, {Ge}, {Lissauer}, {Mann}, {McLean},
  {Mullally}, {Narita}, {Plavchan}, {Ricker}, {Sasselov}, {Seager}, {Sharma},
  {Shiao}, {Sozzetti}, {Stello}, {Vanderspek}, {Wallace}, \&
  {Winn}}]{2019Stassun}
{Stassun}, K.~G., {Oelkers}, R.~J., {Paegert}, M., {et~al.} 2019, \aj, 158, 138

\bibitem[{{Strai{\v{z}}ys} \&
  {Lazauskait{\.{e}}}(2009)}]{2009StraizysLazauskaite}
{Strai{\v{z}}ys}, V. \& {Lazauskait{\.{e}}}, R. 2009, Baltic Astronomy, 18, 19

\bibitem[{{Thompson} {et~al.}(2019){Thompson}, {Kochanek}, {Stanek}, {Badenes},
  {Post}, {Jayasinghe}, {Latham}, {Bieryla}, {Esquerdo}, {Berlind}, {Calkins},
  {Tayar}, {Lindegren}, {Johnson}, {Holoien}, {Auchettl}, \&
  {Covey}}]{2019ThompsonBH}
{Thompson}, T.~A., {Kochanek}, C.~S., {Stanek}, K.~Z., {et~al.} 2019, Science,
  366, 637

\bibitem[{{Tsapras} {et~al.}(2009){Tsapras}, {Street}, {Horne}, {Snodgrass},
  {Dominik}, {Allan}, {Steele}, {Bramich}, {Saunders}, {Rattenbury}, {Mottram},
  {Fraser}, {Clay}, {Burgdorf}, {Bode}, {Lister}, {Hawkins}, {Beaulieu},
  {Fouqu{\'e}}, {Albrow}, {Menzies}, {Cassan}, \& {Dominis-Prester}}]{LCO}
{Tsapras}, Y., {Street}, R., {Horne}, K., {et~al.} 2009, Astronomische
  Nachrichten, 330, 4

\bibitem[{{Udalski} {et~al.}(1992){Udalski}, {Szymanski}, {Kaluzny}, {Kubiak},
  \& {Mateo}}]{1992AcA....42..253U}
{Udalski}, A., {Szymanski}, M., {Kaluzny}, J., {Kubiak}, M., \& {Mateo}, M.
  1992, \actaa, 42, 253

\bibitem[{{Wang} \& {Chen}(2019)}]{redToExtinct}
{Wang}, S. \& {Chen}, X. 2019, \apj, 877, 116

\bibitem[{{Witt} \& {Mao}(1994)}]{1994ApJ...430..505W}
{Witt}, H.~J. \& {Mao}, S. 1994, \apj, 430, 505

\bibitem[{{Wyrzykowski} \& {Hodgkin}(2012)}]{2012IAUS..285..425W}
{Wyrzykowski}, {\L}. \& {Hodgkin}, S. 2012, in IAU Symposium, Vol. 285, New
  Horizons in Time Domain Astronomy, ed. E.~{Griffin}, R.~{Hanisch}, \&
  R.~{Seaman}, 425--428

\bibitem[{{Wyrzykowski} {et~al.}(2016){Wyrzykowski}, {Kostrzewa-Rutkowska},
  {Skowron}, {Rybicki}, {Mr{\'o}z}, {Koz{\l}owski}, {Udalski}, {Szyma{\'n}ski},
  {Pietrzy{\'n}ski}, {Soszy{\'n}ski}, {Ulaczyk}, {Pietrukowicz}, {Poleski},
  {Pawlak}, {I{\l}kiewicz}, \& {Rattenbury}}]{2016WyrzykBH}
{Wyrzykowski}, {\L}., {Kostrzewa-Rutkowska}, Z., {Skowron}, J., {et~al.} 2016,
  \mnras, 458, 3012

\bibitem[{{Wyrzykowski} \& {Mandel}(2020)}]{2020WyrzykowskiMandelBH}
{Wyrzykowski}, {\L}. \& {Mandel}, I. 2020, \aap, 636, A20

\bibitem[{{Wyrzykowski} {et~al.}(2020){Wyrzykowski}, {Mr{\'o}z}, {Rybicki},
  {Gromadzki}, {Ko{\l}aczkowski}, {Zieli{\'n}ski}, {Zieli{\'n}ski},
  {Britavskiy}, {Gomboc}, {Sokolovsky}, {Hodgkin}, {Abe}, {Aldi}, {AlMannaei},
  {Altavilla}, {Al Qasim}, {Anupama}, {Awiphan}, {Bachelet}, {Bak{\i}{\c{s}}},
  {Baker}, {Bartlett}, {Bendjoya}, {Benson}, {Bikmaev}, {Birenbaum},
  {Blagorodnova}, {Blanco-Cuaresma}, {Boeva}, {Bonanos}, {Bozza}, {Bramich},
  {Bruni}, {Burenin}, {Burgaz}, {Butterley}, {Caines}, {Caton}, {Calchi
  Novati}, {Carrasco}, {Cassan}, {{\v{C}}epas}, {Cropper},
  {Chru{\'s}li{\'n}ska}, {Clementini}, {Clerici}, {Conti}, {Conti}, {Cross},
  {Cusano}, {Damljanovic}, {Dapergolas}, {D'Ago}, {de Bruijne}, {Dennefeld},
  {Dhillon}, {Dominik}, {Dziedzic}, {Erece}, {Eselevich}, {Esenoglu}, {Eyer},
  {Figuera Jaimes}, {Fossey}, {Galeev}, {Grebenev}, {Gupta}, {Gutaev},
  {Hallakoun}, {Hamanowicz}, {Han}, {Handzlik}, {Haislip}, {Hanlon}, {Hardy},
  {Harrison}, {van Heerden}, {Hoette}, {Horne}, {Hudec}, {Hundertmark},
  {Ihanec}, {Irtuganov}, {Itoh}, {Iwanek}, {Jovanovic}, {Janulis},
  {Jel{\'\i}nek}, {Jensen}, {Kaczmarek}, {Katz}, {Khamitov}, {Kilic},
  {Klencki}, {Kolb}, {Kopacki}, {Kouprianov}, {Kruszy{\'n}ska}, {Kurowski},
  {Latev}, {Lee}, {Leonini}, {Leto}, {Lewis}, {Li}, {Liakos}, {Littlefair},
  {Lu}, {Manser}, {Mao}, {Maoz}, {Martin-Carrillo}, {Marais},
  {Maskoli{\={u}}nas}, {Maund}, {Meintjes}, {Melnikov}, {Ment},
  {Miko{\l}ajczyk}, {Morrell}, {Mowlavi}, {Mo{\'z}dzierski}, {Murphy},
  {Nazarov}, {Netzel}, {Nesci}, {Ngeow}, {Norton}, {Ofek},
  {Pak{\v{s}}tien{\.{e}}}, {Palaversa}, {Pandey}, {Paraskeva}, {Pawlak},
  {Penny}, {Penprase}, {Piascik}, {Prieto}, {Qvam}, {Ranc},
  {Rebassa-Mansergas}, {Reichart}, {Reig}, {Rhodes}, {Rivet}, {Rixon},
  {Roberts}, {Rosi}, {Russell}, {Zanmar Sanchez}, {Scarpetta}, {Seabroke},
  {Shappee}, {Schmidt}, {Shvartzvald}, {Sitek}, {Skowron}, {{\'S}niegowska},
  {Snodgrass}, {Soares}, {van Soelen}, {Spetsieri},
  {Stankevi{\v{c}}i{\={u}}t{\.{e}}}, {Steele}, {Street}, {Strobl}, {Strubble},
  {Szegedi}, {Tinjaca Ramirez}, {Tomasella}, {Tsapras}, {Vernet}, {Villanueva},
  {Vince}, {Wambsganss}, {van der Westhuizen}, {Wiersema}, {Wium}, {Wilson},
  {Yoldas}, {Zhuchkov}, {Zhukov}, {Zdanavi{\v{c}}ius}, {Zo{\l}a}, \&
  {Zubareva}}]{2020Gaia16aye}
{Wyrzykowski}, {\L}., {Mr{\'o}z}, P., {Rybicki}, K.~A., {et~al.} 2020, \aap,
  633, A98

\bibitem[{{Wyrzykowski} {et~al.}(2015){Wyrzykowski}, {Rynkiewicz}, {Skowron},
  {Koz{\l}owski}, {Udalski}, {Szyma{\'n}ski}, {Kubiak}, {Soszy{\'n}ski},
  {Pietrzy{\'n}ski}, {Poleski}, {Pietrukowicz}, \& {Pawlak}}]{2015Wyrzyk}
{Wyrzykowski}, {\L}., {Rynkiewicz}, A.~E., {Skowron}, J., {et~al.} 2015, \apjs,
  216, 12

\bibitem[{{Zieli{\'n}ski} {et~al.}(2020){Zieli{\'n}ski}, {Wyrzykowski},
  {Miko{\l}ajczyk}, {Rybicki}, \& {Ko{\l}aczkowski}}]{2020CPCS2}
{Zieli{\'n}ski}, P., {Wyrzykowski}, {\l}., {Miko{\l}ajczyk}, P., {Rybicki}, K.,
  \& {Ko{\l}aczkowski}, Z. 2020, in XXXIX Polish Astronomical Society Meeting,
  ed. K.~{Ma{\l}ek}, M.~{Poli{\'n}ska}, A.~{Majczyna}, G.~{Stachowski},
  R.~{Poleski}, {\L}.~{Wyrzykowski}, \& A.~{R{\'o}{\.z}a{\'n}ska}, Vol.~10,
  190--193

\bibitem[{{Zieli{\'n}ski} {et~al.}(2019){Zieli{\'n}ski}, {Wyrzykowski},
  {Rybicki}, {Ko{\l}aczkowski}, {Bru{\'s}}, \& {Miko{\l}ajczyk}}]{2019CPCS2}
{Zieli{\'n}ski}, P., {Wyrzykowski}, {\L}., {Rybicki}, K., {et~al.} 2019,
  Contributions of the Astronomical Observatory Skalnate Pleso, 49, 125

\end{thebibliography}

\begin{appendix}
\section{Photometry}
Here, we present the photometric data used for modelling Gaia18cbf.

\begin{table}
\caption{\label{tab:photGaia}Gaia18cbf photometry from Gaia Science Alerts. The uncertainties (err[mag]) were estimated using Gaia DR2 (see Section \ref{sec:phot}). Full, machine-readable version of this table is available at the CDS.}
\centering
\begin{tabular}{c c c}
\hline
\noalign{\smallskip}
JD-245000. &  $i$ [mag] & err[mag] \\
\noalign{\smallskip}
\hline
\hline
\noalign{\smallskip}
2457032.6801 & 20.280 & 0.103 \\
2457032.7542 & 20.360 & 0.108 \\
2457039.9346 & 20.260 & 0.102 \\
2457040.0086 & 20.130 & 0.094 \\
2457040.1848 & 20.280 & 0.103 \\
2457040.2588 & 20.360 & 0.108 \\
2457127.4686 & 20.270 & 0.102 \\
2457127.6447 & 20.320 & 0.105 \\
2457127.7187 & 20.210 & 0.099 \\
2457127.8949 & 20.290 & 0.104 \\
2457214.4278 & 20.350 & 0.107 \\
2457214.6779 & 20.290 & 0.104 \\
2457214.9281 & 20.340 & 0.107 \\
2457215.0021 & 20.280 & 0.103 \\
2457215.2522 & 20.220 & 0.099 \\
2457215.5024 & 20.280 & 0.103 \\
2457220.0049 & 20.230 & 0.100 \\
2457220.1810 & 20.260 & 0.102 \\
2457220.4312 & 20.210 & 0.099 \\
2457220.5052 & 20.200 & 0.098 \\
2457220.6813 & 20.260 & 0.102 \\
2457262.6578 & 20.310 & 0.105 \\
2457262.7318 & 20.250 & 0.101 \\
2457311.9537 & 20.210 & 0.099 \\
2457405.0450 & 20.200 & 0.098 \\
2457420.4756 & 20.210 & 0.099 \\
2457420.5496 & 20.250 & 0.101 \\
2457454.2045 & 20.310 & 0.105 \\
2457585.9715 & 20.150 & 0.095 \\
2457586.0455 & 20.170 & 0.096 \\
2457604.7991 & 20.140 & 0.095 \\
2457779.0896 & 20.040 & 0.089 \\
2457800.0733 & 20.080 & 0.091 \\
2457800.1473 & 20.140 & 0.095 \\
2457829.8853 & 20.070 & 0.091 \\
2457960.8257 & 19.780 & 0.076 \\
2457960.8998 & 19.730 & 0.074 \\
2457985.8224 & 19.770 & 0.076 \\
2457985.8964 & 19.780 & 0.076 \\
2458014.3832 & 19.790 & 0.077 \\
... & ... & ... \\
\noalign{\smallskip}
\hline
\end{tabular}
\end{table}

\begin{table}
\caption{\label{tab:photGaiaCSPSi}Gaia18cbf photometry from LCO i-band. The machine-readable version of this table is available at the CDS.}
\centering
\begin{tabular}{c c c}
\hline
\noalign{\smallskip}
JD-245000. &  $i$ [mag] & err[mag] \\
\noalign{\smallskip}
\hline
\hline
\noalign{\smallskip}
2458602.2457 & 18.486 & 0.065 \\
2458602.2476 & 18.493 & 0.064 \\
2458740.8956 & 19.240 & 0.065 \\
2458740.8978 & 19.261 & 0.065 \\
2458751.9437 & 19.281 & 0.080 \\
2458752.9173 & 19.285 & 0.071 \\
2458753.9158 & 19.275 & 0.062 \\
2458756.8798 & 19.242 & 0.066 \\
2458756.9224 & 19.256 & 0.069 \\
2458757.9023 & 19.257 & 0.064 \\
2458758.9027 & 19.259 & 0.062 \\
2458926.1310 & 19.408 & 0.066 \\
2458929.2024 & 19.432 & 0.064 \\
2458940.0734 & 19.440 & 0.064 \\
2458956.2838 & 19.598 & 0.075 \\
2458962.0046 & 19.516 & 0.066 \\
2458965.1840 & 19.511 & 0.068 \\
2458968.0059 & 19.574 & 0.065 \\
2458973.9752 & 19.594 & 0.066 \\
2458982.9373 & 19.557 & 0.063 \\
2458983.0010 & 19.557 & 0.065 \\
2458985.9600 & 19.568 & 0.066 \\
2459007.8684 & 19.579 & 0.067 \\
2459011.1633 & 19.667 & 0.141 \\
2459027.1532 & 19.790 & 0.081 \\
2459027.8676 & 19.645 & 0.065 \\
2459028.8499 & 19.615 & 0.069 \\
2459029.8695 & 19.648 & 0.073 \\
2459030.8910 & 19.659 & 0.084 \\
2459035.8586 & 19.686 & 0.122 \\
2459037.0737 & 19.853 & 0.106 \\
2459038.0837 & 19.707 & 0.074 \\
2459038.9353 & 19.666 & 0.062 \\
2459067.0389 & 19.675 & 0.093 \\
\noalign{\smallskip}
\hline
\end{tabular}
\end{table}
\end{appendix}
\end{document}